\documentclass{article}
\usepackage{etex}

\usepackage[reqno]{amsmath}
\usepackage{amssymb}
\usepackage{amsthm}
\usepackage{stmaryrd}
\usepackage{cmll}
\usepackage{nameref,hyperref,cleveref}
\usepackage[normalem]{ulem}
\usepackage[utf8]{inputenc} 
\usepackage[T1]{fontenc}
\usepackage{mathrsfs}
\usepackage{proof}

\usepackage{color}
\usepackage{graphicx}
\usepackage[all,2cell]{xy}
\UseTwocells

\usepackage{tcolorbox}
\tcbset{colback=white,boxrule=0.2mm,boxsep=-0.5mm,top=-1mm,bottom=1mm}

\newtheorem{theorem}{Theorem}[section]
\crefname{theorem}{thm.}{theorems}

\crefname{corollary}{cor.}{corollaries}

\newtheorem{proposition}[theorem]{Proposition}
\crefname{proposition}{prop.}{propositions}

\theoremstyle{definition}
\newtheorem{definition}[theorem]{Definition}
\crefname{definition}{defn.}{definitions}

\PassOptionsToPackage{svgnames}{xcolor}
\usepackage{tikz}
\usetikzlibrary{decorations.markings,snakes,shapes.geometric,calc}

\definecolor{bleu}{RGB}{201,232,251}
\definecolor{rouge}{RGB}{243,153,123}

\tikzstyle{none}=[inner sep=0pt]
\tikzset{->-/.style={decoration={
  markings,
  mark=at position .5 with {\arrow{>}}},postaction={decorate}}
}

\tikzstyle{directed}=[->-,thick]
\tikzstyle{wire}=[thick]
\tikzstyle{rel}=[none,inner sep=2pt]
\tikzstyle{sep}=[thin]
\tikzstyle{ghost}=[dashed,rouge]
\tikzstyle{pt}=[circle,fill=Black,draw=Black,scale=0.3]
\tikzstyle{corner}=[none]

\tikzset{every draw/.append style={thick}}

\pgfdeclarelayer{edgelayer}
\pgfdeclarelayer{nodelayer}
\pgfdeclarelayer{boxlayer}
\pgfdeclarelayer{framelayer}
\pgfsetlayers{framelayer,boxlayer,edgelayer,nodelayer,main}

\newcommand\defeq{\stackrel{\text{\tiny def}}{=}}

\newcommand*{\vcenteredhbox}[1]{\begingroup\setbox0=\hbox{#1}\parbox{\wd0}{\box0}\endgroup}

\newcommand\id{{\rm id}}
\newcommand\op{{\,op}}

\newcommand\Set{ {\mathbf{Set} } }
\newcommand\Cat{ {\mathbf{Cat} } }
\newcommand\Gpd{ {\mathbf{Gpd} } }

\newcommand\Rel{ {\mathbf{Rel} } }

\newcommand\Matplus{ {\mathbf{Mat}_{\oplus} } }
\newcommand\Matminus{ {\mathbf{Mat}_{\ominus} } }

\newcommand\MAT{ {\mathbf{Mat}_{\bullet} } }
\newcommand\coMAT{ {\mathbf{Mat}_{\circ} } }

\newcommand\SubSet{ {\mathbf{SubSet} } }
\newcommand\SUBSET{ {\mathbf{Rel}_{\bullet} } }
\newcommand\coSUBSET{ {\mathbf{Rel}_{\circ} } }
\newcommand\coSUBSETop{ {\mathbf{Rel}_{\circ}^{\,op} } }
\newcommand\Psh{\mathbf{Psh}}
\newcommand\Dist{\mathbf{Dist}}
\newcommand\PSH{ \Dist_\bullet }
\newcommand\coPSH{ \Dist_\circ }

\newcommand\refs{\sqsubset}
\newcommand\seq[1]{\underset{#1}\Longrightarrow}

\newcommand\viso{\equiv}

\newcommand\pullverbatim[1]{\mathbf{pull}_{#1}}
\newcommand\pushverbatim[1]{\mathbf{push}_{#1}}

\newcommand\pullD[1]{\forall_{#1}}
\newcommand\pushD[1]{\exists_{#1}}

\newcommand\ImpR{\mathbin{\text{\reflectbox{$\multimap$}}}}

\newcommand\set[1]{\{\,#1\,\}}

\newcommand\dto{\nrightarrow}

\newcommand{\emb}{\mathrm{emb}}
\newcommand{\embP}[1]{{#1}^{\oplus}}
\newcommand{\embN}[1]{{#1}^{\ominus}}

\newcommand{\morph}[1]{\stackrel{#1}{\longrightarrow}}

\newcommand{\Acategory}{\mathscr{A}}
\newcommand{\Bcategory}{\mathscr{B}}
\newcommand{\Ccategory}{\mathscr{C}}

\newcommand{\Ecategory}{\mathscr{E}}
\newcommand{\Fcategory}{\mathscr{F}}

\newcommand{\Vcategory}{\mathscr{V}}

\newcommand{\predicates}[1]{\mathscr{P}_{#1}}
\newcommand{\substitution}[1]{\mathscr{P}_{#1}}
\newcommand{\powerset}[1]{\mathscr{P}_{#1}}
\newcommand{\qpredicates}[1]{\mathscr{Q}_{#1}}
\newcommand{\predicatesP}[1]{\mathscr{P}^{\oplus}_{#1}}
\newcommand{\predicatesN}[1]{\mathscr{P}^{\ominus}_{#1}}
\newcommand{\predicatesRel}[1]{\mathscr{R}_{#1}}

\newcommand{\lawvereidentity}[1]{\mathbf{I}_{#1}}
\newcommand{\relationalidentity}[1]{\mathbf{J}_{#1}}

\newcommand{\tensorialand}{\varowedge}
\newcommand{\tensorialor}{\varovee}
\newcommand{\tensorialtrue}{\mathbf{true}}
\newcommand{\tensorialfalse}{\mathbf{false}}

\newcommand{\timesSet}{\times_{\Set}}
\newcommand{\impliesSet}{\to_{\Set}}

\newcommand{\graphD}[1]{\langle{#1}\rangle}

\begin{document}
\title{A bifibrational reconstruction of \\ Lawvere's presheaf hyperdoctrine}
\author{Paul-Andr\'e Melli\`es\footnote{Institut de Recherche en Informatique Fondamentale, CNRS, Universit\'e Paris Diderot, Sorbonne Paris Cit\'e.} \and Noam Zeilberger\footnote{Inria, Équipe Parsifal, Bâtiment Alan Turing, Campus de l'École Polytechnique.}}

\maketitle

\begin{abstract}
Combining insights from the study of type refinement systems and of
monoidal closed chiralities, we show how to reconstruct Lawvere's
hyperdoctrine of presheaves using a full and faithful embedding into a
monoidal closed bifibration living now over the compact closed
category of small categories and distributors.  Besides revealing
dualities which are not immediately apparent in the traditional
presentation of the presheaf hyperdoctrine, this reconstruction leads
us to an axiomatic treatment of directed equality predicates (modelled
by $\hom$ presheaves), realizing a vision initially set out by Lawvere
(1970).  It also leads to a simple calculus of string diagrams
(representing presheaves) that is highly reminiscent of C.~S.~Peirce's
existential graphs for predicate logic, refining an earlier
interpretation of existential graphs in terms of Boolean
hyperdoctrines by Brady and Trimble.  Finally, we illustrate how this
work extends to a bifibrational setting a number of fundamental ideas
of linear logic.
\end{abstract}

\section{Introduction}
\label{section/intro}

\paragraph{An intriguing discrepancy.}
There is an intriguing and longrunning discrepancy in categorical logic between
the way conjunction is coupled to implication in cartesian closed categories,
and the way existential quantification is coupled to universal quantification
in hyperdoctrines.
In a cartesian closed category~$\Ccategory$, every object~$A$
induces an adjunction
\begin{equation}\label{equation/ccc-adjunction}
A\times - \quad \dashv \quad A\Rightarrow - 
\end{equation}
where the implication functor
$$B\mapsto A\Rightarrow B
\quad : \quad \Ccategory \quad \morph{} \quad \Ccategory$$
is right adjoint to the conjunction functor
$$B\mapsto A\times B \quad : \quad \Ccategory \quad \morph{} \quad \Ccategory.$$ 
This categorical situation should be compared with the way quantification is handled in a hyperdoctrine.
Recall that a hyperdoctrine in the sense of Lawvere is first of all
a (pseudo) functor
$$\predicates{} \quad : \quad \Bcategory^{\,op} \quad \morph{} \quad \Cat$$
from a base category~$\Bcategory$ to the category $\Cat$ of small categories and functors.
The intuition behind this definition is that every object~$A$ of the category~$\Bcategory$ 
is assigned a ``category of predicates'' noted~$\predicates{A}$, and every morphism $f:A\to B$ of $\Bcategory$ induces a functor
$$
\substitution{f} \quad : \quad \predicates{B} \quad \morph{} \quad  \predicates{A}
$$
called ``substitution'' along~$f$.
The leading example of a hyperdoctrine is the ``subset hyperdoctrine''
with basis the category $\Bcategory=\Set$ of sets and functions,
equipped with the powerset functor~$\predicates{}$ which transports
every set $A$ to the set~$(\powerset{A},\subseteq)$ of subsets of~$A$
ordered by inclusion.
Note that the ordered set $\powerset{A}$ is seen here as the ordered category
where two subsets $R,S\subseteq A$ are related by a morphism $R\to S$
precisely when $R\subseteq S$.
The substitution functor along a function $f:A\to B$
is defined by transporting every subset $S\subseteq B$ to its inverse image
$$
\substitution{f} \quad = \quad S \quad \mapsto \quad
\{ \hspace{.5em} a\in A \hspace{.5em} | \hspace{.5em} fa \in S \hspace{.5em} \}.
$$
The definition of a hyperdoctrine then additionally asks for a pair of functors
$$\Sigma_f \, , \, \Pi_f \quad : \quad \substitution{A} \quad \morph{} \quad \substitution{B}$$  
called ``existential quantification'' and ``universal quantification'' along $f$, which are respectively left and right adjoint to the substitution functor:
\begin{equation}\label{equation/hyperdoctrine-adjunction}
\Sigma_f \quad \dashv \quad \substitution{f} \quad \dashv \quad \Pi_f
\end{equation}
In the case of the subset hyperdoctrine,
the functors $\Sigma_f$ and $\Pi_f$ transport a subset~$R\subseteq A$
to the following subsets of~$B$:
$$
\begin{array}{ccc}
\Sigma_f \, R  &  =  & \{ \hspace{.5em} b\in B \hspace{.5em} | \hspace{.5em} \exists a\in A,  \hspace{.5em}  fa=b \hspace{.5em} \wedge \hspace{.5em} a\in R \hspace{.5em} \}
\\
\Pi_f \, R  &  =  & \{ \hspace{.5em} b\in B \hspace{.5em} | \hspace{.5em} \forall a\in A, \hspace{.5em} fa=b \hspace{.35em}\Rightarrow \hspace{.3em} a\in R \hspace{.5em} \}
\end{array}
$$
The difference between (\ref{equation/ccc-adjunction}) and (\ref{equation/hyperdoctrine-adjunction})
is especially notable if one thinks of dependent type theory, where existential quantification 
provides a dependent form of conjunction, and universal quantification a dependent form of implication.
It is thus puzzling to see conjunction and implication directly coupled by an adjunction in~(\ref{equation/ccc-adjunction})
while they form in~(\ref{equation/hyperdoctrine-adjunction})
a ``m\'enage \`a trois'' with the substitution functor~$\predicates{f}$ as intermediate.

\medbreak

In the present introduction, we explain how to reconcile the two points of view 
in the specific subset hyperdoctrine on $\Bcategory=\Set$.
The choice of this hyperdoctrine is mainly pedagogical:
we find clarifying to explain some of our ideas in this familiar example.
However, as we will see, the ideas developed in this introduction lift very smoothly
to the more sophisticated situation when one replaces $\Bcategory=\Set$
by the cartesian closed category $\Bcategory=\Cat$
of small categories and functors, and where the ``category of predicates'' $\predicates{A}$
over a small category~$A$ is defined as the contravariant presheaf category
$$\predicates{A} \quad := \quad \hat{A} \quad \defeq \quad [A^\op,\Set].$$
Lawvere introduced this example in his original article on hyperdoctrines \cite{lawvere69}, and also considered its restriction to presheaves over groupoids ($\Bcategory = \Gpd$) in his article describing a treatment of equality in hyperdoctrines, which relied on a ``Frobenius Reciprocity'' condition and certain Beck-Chevalley conditions \cite{lawvere70}.
The presheaf hyperdoctrine is an important example despite the fact that it does not in general satisfy these conditions, and indeed, Lawvere even writes that this fact <<\,should not be taken as indicative of a lack of vitality
[...] or even of a lack of a satisfactory theory of equality\,>> for the presheaf hyperdoctrine, but rather <<\,that we have probably been too naive in defining equality 
in a manner too closely suggested by the classical conception\,>> \cite[p.11]{lawvere70}. 
We will come back to this important point later in the introduction.
\paragraph{From functions to relations.}
Our procedure to reconcile (\ref{equation/ccc-adjunction}) and (\ref{equation/hyperdoctrine-adjunction})
is inspired by linear logic and the shift from the cartesian closed category~$\Set$
to the symmetric monoidal closed (and in fact, compact closed) category~$\Rel$ which underlies its discovery by Girard~\cite{girard-linear-logic}.
In particular, we will make a great usage of the two ``embedding'' functors
$$
\embP{\emb}:\Set \to \Rel
\quad\quad\quad
\embN{\emb}:\Set^\op \to \Rel
$$
which transport a set~$A$ to itself, and a function~$f:A\to B$
to the binary relations
$$
\embP{f}:A\dto B
\quad\quad\quad
\embN{f}:B\dto A
$$
where
$$
\begin{array}{ccc}
\embP{f} & = & \set{(a,b)\in A\times B \mid  fa=b \, }
\\
\embN{f} & = & \set{(b,a)\in B\times A \mid \, b=fa }
\end{array}
$$
Notation: we write $M:A\dto B$ for a binary relation $M\subseteq A\times B$
which defines a morphism $A\to B$ in the category~$\Rel$.
These two faithful (but not full) functors $\embP{\emb}$ and $\embN{\emb}$
transport the category~$\Set$ and its opposite category~$\Set^\op$
in the same category~$\Rel$.

\paragraph{Bifibrations.}
Another important ingredient and source of inspiration for our work 
is the notion of \emph{bifibration} which we like to see as a particular instance
of \emph{type refinement system} in the terminology of~\cite{mz15popl,mz15isbell}.
A bifibration may be defined as a functor
$$
p \quad : \quad \Ecategory \quad \morph{} \quad \Bcategory
$$
which is at the same time a fibration and an opfibration.
Following the principles and notations of type refinement systems, an object $R$ of $\Ecategory$ is said to ``refine'' an object $A$ of $\Bcategory$ (written $R \refs A$) if $p(R) = A$, while a ``derivation'' of a typing judgment
$$R \seq{f} S$$
(where $f : A \to B$, $R \refs A$, and $S \refs B$) is defined as a morphism $\alpha : R\to S$ in the category~$\Ecategory$ whose image by the functor~$p$
is the morphism~$f$.
The definition of a bifibration then asserts that the functor (or ``refinement system'') $p$ is equipped with operations for pushing or pulling an object of $\Ecategory$ along a morphism of $\Bcategory$
$$
\infer{\pushverbatim{f}R \refs B}{R \refs A & f : A \to B}
\qquad
\infer{\pullverbatim{f}S \refs A}{f : A \to B & S \refs B}
$$
such that there is a one-to-one correspondence of derivations,
$$
\infer={\pushverbatim{f}R \seq{g} R'}{R \seq{f;g} R'}
\qquad
\infer={S' \seq{e} \pullverbatim{f}S}{S' \seq{e;f} S}
$$
written here as invertible inference rules in the proof-theoretic style of refinement systems (see \cite{mz15popl,mz15isbell} for details).
Notably, this structure is sufficient to derive inference rules
$$
\infer{\pushverbatim{f}R_1 \seq{\id_B} \pushverbatim{f}R_2}{R_1 \seq{\id_A} R_2}
\qquad
\infer{\pullverbatim{f}S_1 \seq{\id_A} \pullverbatim{f}S_2}{S_1 \seq{\id_B} S_2}
$$
as well as isomorphisms
$$
\pullverbatim{(f;g)}S \viso \pullverbatim{f}\pullverbatim{g}S
\qquad
\pullverbatim{\id}S \viso S
$$
$$
\pushverbatim{(g\circ f)}R \viso \pushverbatim{g}\pushverbatim{f}R
\qquad
\pushverbatim{\id}R \viso R
$$
and a three-way correspondence of derivations:
$$
\infer={R\seq{\id_A} \pullverbatim{f}(S)}{
\infer={R \seq{f} S}{
\pushverbatim{f} (R)\seq{\id_B} S}}
$$
This argument establishes that any (cloven) bifibration $p : \Ecategory \to \Bcategory$ determines a pair of (pseudo) functors
$$\pushverbatim{} : \Bcategory \morph{} \Cat \qquad
\pullverbatim{} : \Bcategory^\op \morph{} \Cat$$
as well as a family of adjunctions
$$
\xymatrix{\Ecategory_A \rrtwocell^{\pushverbatim{f}}_{\pullverbatim{f}}{'\bot} && \Ecategory_B}
$$
relating the corresponding functors between the ``fibre'' categories $\Ecategory_A$ and $\Ecategory_B$ (defined as subcategories of $\Ecategory$ containing only those morphisms that project by $p$ to identity morphisms in $\Bcategory$).

\paragraph{A subset bifibration over sets and relations.}
Putting these two sources of inspiration together: linear logic and bifibrations,
we construct a ``subset bifibration''
$$p \quad : \quad \SUBSET \quad \morph{} \quad \Rel$$
where the category $\SUBSET$ has objects the pairs $(A,R)$
consisting of a set~$A$ together with a subset~$R\subseteq A$ ;
and morphisms
$$
M \quad : \quad (A,R)  \to (B,S)
$$
defined as the binary relations $M:A\dto B$ satisfying the property
$$\forall a\in A, \forall b\in B, \quad (\, M(a,b) \, \wedge \, Ra \,) \, \Rightarrow \, Sb.$$
The functor~$p$ transports every object~$(A,R)$ to the first component~$A$,
and every morphism~$M:(A,R)  \to (B,S)$ to the underlying relation $M:A\dto B$.
The category $\SUBSET$ may be seen as a category of ``pointed objects'' in $\Rel$, 
since an object~$(A,R)$ is the same thing as a relation $R:1\dto A$,
with morphisms defined using the 2-categorical structure of $\Rel$ provided by inclusion of binary relations.
The fiber or category of predicates~$\powerset{A}=p^{-1}(A)$ associated to a set~$A$
by the functor~$p$ is simply the set ${(\powerset{A},\subseteq)}$ of subsets of~$A$ ordered by inclusion.
An important point is that the functor~$p$ just defined is a bifibration.
Given a binary relation
$$
M \quad : \quad A \quad \dto \quad B
$$
the two functors
$$
\begin{array}{ccccccc}
\exists_M & = & \pushverbatim{M} & : & \predicates{A} & \morph{} & \predicates{B}
\\
\forall_M & = & \pullverbatim{M} & : & \predicates{B} & \morph{} & \predicates{A}
\end{array}
$$
are defined in the following way:
$$
\begin{array}{ccl}
\exists_M \, R & = & \set{ b\in B \mid \exists a\in A, \, \, M(a,b) \wedge Ra}
\\
\forall_M \, S  & = & \set{ a\in A \mid \forall b\in B, \,\, M(a,b) \Rightarrow Sb}
\end{array}
$$
for all subsets $R\subseteq A$ and $S\subseteq B$.
An easy computation shows that $\exists_M$ and $\forall_M$ define a pair of adjoint functors
\begin{equation}\label{equation/rel-adjunction}
\exists_M \quad \dashv \quad \forall_M
\end{equation}
because
$\exists_M R \, \subseteq S$ is equivalent to $R \subseteq\,  \forall_M S$
for every $R\subseteq A$ and $S\subseteq B$.
From this, we conclude that
\begin{theorem}
The functor $p:\SUBSET\to\Rel$ is a bifibration.
\end{theorem}
\noindent
The associated fibre functor
\begin{equation}\label{equation/subset-bifibration}
\predicatesRel{} \quad : \quad \Rel^{\,op} \quad \morph{} \quad \Cat
\end{equation}
transports every set~$A$ to the set ${(\powerset{A},\subseteq)}$ of subsets of~$A$ ordered by inclusion.

\paragraph{One hyperdoctrine decomposed into two bifibrations.}
The construction of the subset bifibration~$p:\SUBSET\to\Rel$
on sets and relations leads us to a new way to think about existential and universal quantification
in the subset hyperdoctrine~$\powerset{}$ on the category $\Bcategory=\Set$.
Indeed, given a hyperdoctrine 
$$\predicates{} \quad : \quad \Bcategory^{\,op} \quad \morph{} \quad \Cat$$
it is always possible to ``decorrelate'' the pair of adjunctions~(\ref{equation/hyperdoctrine-adjunction})
by defining a pair of (pseudo) functors
$$
\predicatesP{} : \Bcategory^{\,op}\morph{}\Cat
\quad\quad\quad
\predicatesN{} : \Bcategory\morph{}\Cat
$$
where $\predicatesP{}=\predicates{}$ and where $\predicatesN{}$ transports
every object~$A\in\Bcategory$ to the category $\predicates{A}$
and every morphism~$f:A\to B$ to the functor
$$
\predicatesN{f} := \Pi_f \quad : \quad \predicatesN{A} \quad \morph{} \quad \predicatesN{B}.
$$
The key observation here is that the left-hand side adjunction 
$$\Sigma_f \quad \dashv \quad \predicates{f}$$
of the hyperdoctrine~$p$ ensures that $\predicatesP{}$ determines a bifibration with basis the category~$\Bcategory$, 
while the right-hand side adjunction 
$$\predicates{f} \quad \dashv \quad \Pi_f$$
ensures that $\predicatesN{}$ determines a bifibration
with basis the opposite category~$\Bcategory^{\,op}$.
In the case of the subset hyperdoctrine on $\Bcategory=\Set$, one obtains in this way two (pseudo) functors
$$
\predicatesP{} : \Set^{\,op}\morph{}\Cat
\quad\quad\quad
\predicatesN{} : \Set\morph{}\Cat
$$
which provide an alternative and equivalent formulation of the original subset hyperdoctrine~$\powerset{}$
on the category~$\Set$.
In particular, $\predicatesP{}$ and $\predicatesN{}$ determine a pair of bifibrations
$$
\embP{p}:\embP{\SubSet}\to\Set
\quad\quad\quad
\embN{p}:\embN{\SubSet}\to\Set^{\,op}
$$
where the categories $\embP{\SubSet}$ and $\embN{\SubSet}$ have the same objects
defined as pairs $(A,R)$ consisting of a set~$A$ and of a subset~$R\subseteq A$, while the morphisms 
$$f \quad : \quad (A,R) \quad \morph{} \quad (B,S)$$
are defined as the functions $f:A\to B$ satisfying the property
$$
\forall a\in A, \quad Ra \, \Rightarrow \, S(fa)
$$
in the case of $\embP{\SubSet}$ 
and as the functions $f:B\to A$ satisfying the property
$$
\forall b\in B, \quad R(fb) \, \Rightarrow \, Sb
$$
in the case of $\embN{\SubSet}$.
Note the change of orientation in the definition of the morphisms
of $\embP{\SubSet}$ and of $\embN{\SubSet}$.
As expected, the functors~$\embP{p}$ and $\embN{p}$ transport every such 
morphism~$f: (A,R)\to (B,S)$ to the underlying morphism $f:A\to B$
in the category $\Set$ for the functor~$\embP{p}$ and in the category $\Set^\op$ for the functor~$\embN{p}$.

\paragraph{Putting everything back together.}
A quite extraordinary and instructive phenomenon appears at this point: 
the two bifibrations~$\embP{p}$ and $\embN{p}$ 
and thus the hyperdoctrine~$\predicates{}$ on $\Set$ may be recovered 
from the bifibration~$p:\SUBSET\to\Rel$ and the two embedding functors:
$$
\embP{\emb}:\Set \to \Rel
\quad\quad\quad
\embN{\emb}:\Set^\op \to \Rel.
$$
The reason is that, for every function $f:A\to B$, the following equations hold:
$$
\begin{array}{ccl}
\exists_{\embP{f}} \, R & = & \set{ b\in B \mid \exists a\in A, \, \, \embP{f}(a,b) \wedge Ra}
\\
\forall_{\embN{f}} \, R  & = & \set{ b\in B \mid \forall a\in A, \,\, \embN{f}(a,b) \Rightarrow Ra}
\end{array}
$$
for all $R\subseteq A$.
From this follows that
$$
\Sigma_f \, = \, \exists_{\embP{f}}  \quad\quad\quad \Pi_f \, = \, \forall_{\embN{f}}
$$
By uniqueness of a left or of a right adjoint, these two equations 
together with (\ref{equation/hyperdoctrine-adjunction})
and (\ref{equation/rel-adjunction}) imply the series of equalities:
$$
\forall_{\embP{f}}  \, = \, \predicates{f} \, = \, \exists_{\embN{f}}.
$$
The resulting picture reconciles (\ref{equation/ccc-adjunction}) and (\ref{equation/hyperdoctrine-adjunction})
since the original series of adjunctions of a hyperdoctrine~(\ref{equation/hyperdoctrine-adjunction})
is replaced by a pair of adjunctions
$$
\exists_{\embP{f}} \quad \dashv \quad \forall_{\embP{f}}  \quad = \quad \exists_{\embN{f}} \quad \dashv \quad\forall_{\embN{f}}
$$
living in two different bifibrations $\embP{p}$ and $\embN{p}$, 
together with an equality between the two functors $\forall_{\embP{f}}$ and $\exists_{\embN{f}}$.
An interesting outcome of our decomposition of the subset hyperdoctrine~$\predicates{}$
over $\Set$ is that the existential quantification is entirely handled by the bifibration~$\embP{p}$
while the universal quantification is entirely handled by the bifibration~$\embN{p}$.
The decomposition reveals moreover
that the substitution functor $\predicates{f}$ of the subset hyperdoctrine
is not primitive, since it is the ``superposition'' of the two equal functors $\forall_{\embP{f}}$ and $\exists_{\embN{f}}$.

\medbreak

Formally speaking, recall that every bifibration $p:\Ecategory\to\Bcategory$ may be ``pulled back'' along a functor $F:\Ccategory\to\Bcategory$
in order to define a bifibration $q:\Fcategory\to\Ccategory$ on the category~$\Ccategory$:
$$
\xymatrix @-1.9pc {
\Fcategory
\ar[rrrrrrrrrr]^{}
\ar[dddd]_-{q}
\ar@{}[ddddrrrrrrrrrr]|-{pullback}
&&& &&&&&&&
\Ecategory
\ar[dddd]^{p}
\\
\\
&&& &&&&&&&
\\
\\
\Ccategory \ar[rrrrrrrrrr]^{F} &&&&&&&&&& \Bcategory
}
$$
whose fibre functor~$\qpredicates{}$ is simply obtained by precomposing 
the fibre functor~$\predicates{}$ of the bifibration~$p$ with the functor~$F^{\,op}$:
$$
\qpredicates{} = \predicates{}\circ F^{\,op} \quad : \quad \Ccategory^{\,op} \quad \morph{F^\op} \quad \Bcategory^{\,op} \quad \morph{} \quad \Cat.
$$
In other words, the category of predicates~$\qpredicates{A}$ associated to an object~$A$ of the category~$\Ccategory$
coincides with the category of predicates~$\predicates{FA}$ of its image by the functor~$F$.
In the discussion above, we have just established that 
\begin{theorem}\label{thm/reconstruction}
The two bifibrations
$$
\embP{p} : \embP{\SubSet}\morph{}\Set
\quad\quad\quad
\embN{p}  : \embN{\SubSet}\morph{}\Set^\op
$$
are equal to the bifibration
$$p \quad : \quad \SUBSET \quad \morph{} \quad \Rel$$
pulled back along the embedding functors $\embP{\emb}:\Set \to \Rel$
and $\embN{\emb}:\Set^\op \to \Rel$.
\end{theorem}
\noindent
This means that there exists a pair of pullback diagrams
$$
\xymatrix @-1.9pc {
\embP{\SubSet}
\ar[rrrrrr]
\ar[dddd]_-{\embP{p}}
\ar@{}[ddddrrrrrr]|-{pullback}
&&&&&& 
\SUBSET
\ar@{}[ddddrrrrrr]|-{pullback}
\ar[dddd]^{p}
&&&&&&
\embN{\SubSet}
\ar[llllll]
\ar[dddd]^-{\embN{p}}
\\
\\
&& &&&&&& && &&
\\
\\
\Set \ar[rrrrrr]^-{\embP{\emb}} 
&&&& &&
\Rel
&&&&&&
\Set^{\,op} 
\ar[llllll]_-{\embN{\emb}} 
}
$$
which enable us to derive the two bifibrations~$\embP{p}$ and~$\embN{p}$
and thus the subset hyperdoctrine~$\powerset{}$ on the category~$\Set$ of sets and functions,
from the subset bifibration~$p$ on the category~$\Rel$ of sets and relations.

\paragraph{A monoidal closed refinement system.}
Now that we have given theoretical precedence to the subset bifibration $p:\SUBSET\to\Rel$
over the hyperdoctrine $\powerset{}$ on $\Set$, there remains to study the properties
of this bifibration~$p$ more closely.
In our work on refinement systems, we have advocated the fundamental role
played by the interaction between the adjunctions of a monoidal closed refinement system
which would be a bifibration at the same time:
$$
R\otimes - \dashv R\multimap -
\quad\quad\quad
\pushverbatim{f}\dashv\pullverbatim{f}
$$
where $R$ is a refinement and $f:A\to B$ a morphism of the basis category.
Recall that by ``(symmetric) monoidal closed refinement system'',
we simply mean a functor $p:\Ecategory\to\Bcategory$
where the categories~$\Ecategory$ and $\Bcategory$ are (symmetric) monoidal closed
and where the functor~$p$ preserves the (symmetric) monoidal closed structure of~$\Ecategory$
up to coherent isomorphisms.
A primary observation is that
\begin{theorem}
The refinement system $p : \SUBSET  \morph{}  \Rel{}$ is symmetric monoidal closed
with tensor product and implication of the category~$\SUBSET$ defined as
$$
\begin{array}{ccl}
(A,R)\otimes (B,S) & \defeq & (A\times B,R\otimes S)\\
(A,R)\multimap (B,S) & \defeq & (A\times B,R\multimap S)\\
\end{array}
$$
where the subsets $R\otimes S$ and $R\multimap S$ of $A\times B$ are defined as
$$\begin{array}{ccl}
R\otimes S & = & \set{(a,b)\in A\times B \mid Ra \wedge Sb}
\\
R\multimap S & = & \set{(a,b)\in A\times B \mid Ra \Rightarrow Sb}
\end{array}$$
\end{theorem}
\noindent
Note that the implication $(A,R)\multimap (B,S)$ in $\SUBSET$ is transported
by the functor~$p$ to the set $A\otimes B = A\times B$ which plays the role of internal hom $A\multimap B$
in the compact closed category $(\Rel,\times,1)$.
It is very instructive to study how the adjunctions
$$
R\otimes - \dashv R\multimap - 
\quad\quad\quad
\exists_M \dashv \forall_M 
$$
coming from the bifibrational and monoidal closed structure of~$p$ interact,
for $R=(A,R)$ an object of~$\SUBSET$ and $M$ a morphism of~$\Rel$.
For instance, one has the equality
\begin{equation}\label{equation/exists-vs-tensor}
\exists_{M\otimes N} \,  (R\otimes S) \, \quad = \quad \exists_M R \otimes \exists_N S 
\end{equation}
for all subsets~$R\subseteq A$ and $S\subseteq B$
and relations $M:A\dto C$, $N:B\dto D$.
On the other hand, the canonical inclusion
\begin{equation}\label{equation/inclusion-not-equality}
\forall_M R \otimes \forall_N S 
\quad \subseteq \quad 
\forall_{M\otimes N} \,  (R\otimes S)
\end{equation}
is not an equality in general, for subsets~$R\subseteq A$ and $S\subseteq B$
and relations $M:C\dto A$, $N:D\dto B$.
Consider for instance the case where $C=D=1$,
where the two subsets
$$
\begin{array}{ccc}
\forall_M R \otimes \forall_N S & = & \set{ (a,b) \mid (Ma \Rightarrow Ra)\wedge(Nb \Rightarrow Sb) }
\\
\forall_{M\otimes N} \,  (R\otimes S) & = & \set{ (a,b)\mid (Ma\wedge Nb) \, \Rightarrow \, (Ra \wedge Sb)}
\end{array}
$$
are not equal for general subsets $M,R\subseteq A$ and $N,S\subseteq B$.

\paragraph{Monoidal closed categories as chiralities.}
The observation that $\SUBSET$ is a symmetric monoidal closed category
leads us to the idea of reformulating it as a ``symmetric monoidal closed chirality'' 
in the sense of~\cite{mellies-prims}.
Recall that:
\begin{definition}
A symmetric monoidal closed chirality $(\Acategory,\Bcategory)$
is a pair of symmetric monoidal categories
$$(\Acategory,\tensorialand,\tensorialtrue)
\quad\quad\quad
(\Bcategory,\tensorialor,\tensorialfalse)$$
equipped with a symmetric monoidal equivalence
$$\xymatrix @-1.8pc {
(\Acategory,\tensorialand,\tensorialtrue)
&\ar@<.5ex>[rrrrrrrr]^-{(-)^{\ast}}
&&&&
&&&&
\ar@<.5ex>[llllllll]^-{^{\ast}(-)}
&
(\Bcategory,\tensorialor,\tensorialfalse)^{\,op(0,1)}}
$$
where the exponent $op(0,1)$ means that the orientation of the tensor product~$\tensorialor$
(of dimension~0) and of the morphisms (of dimension~1) have been reversed ; together with two (pseudo)actions
$$
\begin{array}{ccccc}
\tensorialor & : & \Bcategory\times\Acategory & \morph{} & \Acategory
\\
\tensorialand & : & \Acategory\times\Bcategory & \morph{} & \Bcategory
\end{array}
$$
together with two natural bijections:
$$
\begin{array}{ccc}
\Acategory\,(\, m \, \tensorialand \, a_1 \, , \, a_2 \,) & \cong & \Acategory\,(\, a_1 \, , \, m^{\ast}\tensorialor \, a_2 \,)
\\
\Bcategory\,(\,\, {}^\ast n \,\, \tensorialand \, b_1 \, , \,\, b_2\,) & \cong & \Bcategory\,(\,b_1 \, , \,\, n \, \tensorialor \, b_2 \,)
\end{array}
$$
for $m,a_1,a_2\in\Acategory$ and $n,b_1,b_2\in\Bcategory$, satisfying moreover two coherence diagram, 
see \cite{mellies-prims} for details.
\end{definition}

\medbreak

Every symmetric monoidal closed category $(\Ccategory,\otimes,I)$
may be equivalently formulated as the symmetric monoidal closed chirality
defined by the pair of opposite categories:
$$
(\Acategory,\tensorialand,\tensorialtrue)=(\Ccategory,\otimes,I)
\quad
(\Bcategory,\tensorialor,\tensorialfalse)=(\Ccategory,\otimes,I)^{\,op(0,1)}.
$$
The advantage of this formulation is that the intuitionistic implication
of the monoidal closed category $\Acategory=\Ccategory$
may be ``decomposed'' in just the same way as in classical logic or in linear logic:
\begin{equation}\label{equation/chiral-decomposition}
a_1\multimap a_2 \quad := \quad a_1^{\ast}\tensorialor a_2
\end{equation}
where the operation $(-)^{\ast}$ implements an involutive negation,
and where the notation $\tensorialor$ reflects the fact that the tensor product
of~$\Bcategory$ should be understood as a disjunction.

\paragraph{Bifibrations as chiralities.}
One main contribution of the paper is to observe that
the notion of ``chirality'' may be very elegantly adapted 
to the notion of bifibration.
\begin{definition}
A bifibration chirality~$(p,q)$ is a pair of opfibrations~$p$ and~$q$
$$
p: \Ecategory \to \Bcategory
\quad\quad\quad
q: \Fcategory \to \Ccategory
$$
together with a pair of equivalences
$$\xymatrix @-1.9pc {
\Ecategory
&\ar@<.5ex>[rrrrrr]^-{(-)^{\ast}}
&&&&&&
\ar@<.5ex>[llllll]^-{^{\ast}(-)}
&
\Fcategory^\op}
\quad\quad\quad
\xymatrix @-1.9pc {
\Bcategory
&\ar@<.5ex>[rrrrrr]^-{(-)^{\ast}}
&&&&&&
\ar@<.5ex>[llllll]^-{^{\ast}(-)}
&
\Ccategory^\op}
$$
inducing an equivalence of refinement systems:
$$
\xymatrix @-1.9pc {
\Ecategory\ar[ddddd]_-{p}
\ar@<.5ex>[rrrrrrrr]^-{(-)^{\ast}}
&&&&&&&&
\ar@<.5ex>[llllllll]^-{^{\ast}(-)}
\Fcategory^\op\ar[ddddd]^-{q^{\,op}}
\\
\\
\\
\\
\\
\Bcategory
\ar@<.5ex>[rrrrrrrr]^-{(-)^{\ast}}
&&&&&&&&
\ar@<.5ex>[llllllll]^-{^{\ast}(-)}
\Ccategory^\op
}
$$
\end{definition}
\noindent
In a bifibration chirality~$(p,q)$, the opfibration $p$ is automatically a fibration,
where the pullback $\pullverbatim{f}$ can be computed as
\begin{equation}
\pullverbatim{f}S \viso {}^*(\pushverbatim{f^*}S^*) \label{equation/notpushnot-intro}
\end{equation}
Equation (\ref{equation/notpushnot-intro}) follows from the fact that the pushforward $\pushverbatim{f^{\ast}}$
in the opfibration $q$ is a pullback in the fibration $q^\op$, and that equivalences of refinement systems preserve pullbacks. 
We can also derive it more explicitly in proof-theoretic style, from the invertible inferences
$$
\infer={R \seq{g} {}^*(\pushverbatim{f^*}S^*)}{
\infer={\pushverbatim{f^*}S^* \seq{g^*} R^*}{
\infer={S^* \seq{f^*;g^*} R^*}{
R \seq{g;f} S
}}}
$$
The subset bifibration $p:\SUBSET \to \Rel$ can be formulated as a bifibration chirality~$(p,q)$ as follows.
Define $\coSUBSET$ to be the category whose objects are subsets $R\subseteq A$, $S\subseteq B$
and whose morphisms $R \to S$ are binary relations $M : A \dto B$ satisfying the property
$$
\forall (a,b)\in A\times B, \quad\quad M(a,b) \wedge Sb \quad \Rightarrow \quad Ra.
$$
The category $\coSUBSET$ comes equipped with an evident forgetful functor~$q:\coSUBSET\to\Rel$
which defines an opfibration.
We obtain in this way the bifibration chirality:
$$
\xymatrix @-1.9pc {
\SUBSET\ar[ddddd]_-{p}
\ar@<.5ex>[rrrrrrrr]^-{(-)^{\ast}}
&&&&&&&&
\ar@<.5ex>[llllllll]^-{^{\ast}(-)}
\coSUBSETop\ar[ddddd]^-{q^{\,op}}
\\
\\
\\
\\
\\
\Rel
\ar@<.5ex>[rrrrrrrr]^-{(-)^{\ast}}
&&&&&&&&
\ar@<.5ex>[llllllll]^-{^{\ast}(-)}
\Rel^\op
}
$$
where the functor $(-)^{\ast}$ transports a set~$A$ and a subset $R\subseteq A$ to themselves,
and reverses a binary relation~$M:A\dto B$ in the expected way:
$$
M^{\ast} \,\,\, \defeq \,\,\, \set{ (b,a)\in B\times A \mid M(a,b) } \,\,  :  \,\, B\dto A
$$
and similarly for $^{\ast}(-)$. One obtains the equations
\begin{tcolorbox}
\begin{align}
\forall_{M} \, S \quad & \viso \quad {}^{\ast}(\,\exists_{M^*}\,S^*)  \tag{$a$}\label{eqn:a}
\\
\forall_{M} \, S \quad & \viso \quad (\,\exists_{{}^*\!M}\,\,^*\!S\,\,)^{\ast} \tag{$b$}\label{eqn:b}
\end{align}
\end{tcolorbox}
\noindent
where $M:A\dto B$ in both equations and $S\refs B$ in $p$ for equation~(\ref{eqn:a}),
while $S\refs B$ in $q$ for equation~(\ref{eqn:b}).
Note that the universal quantifier~$\forall_M$ of equation~(\ref{eqn:a})
is computed in~$p$ while the universal quantifier~$\forall_M$
of equation~(\ref{eqn:b}) is computed in~$q$.

\medbreak

Not only that, the category $\coSUBSET$ defines together with $\SUBSET$
a symmetric monoidal chirality
$$
\Acategory=(\SUBSET,\tensorialand,\tensorialtrue)
\quad\quad
\Bcategory=(\coSUBSET,\tensorialor,\tensorialfalse).
$$
Putting together the bifibration chirality $(p,q)$ with the symmetric monoidal closed chirality~$(\Acategory,\Bcategory)$,
we may for instance rewrite equation~(\ref{equation/exists-vs-tensor}) as the following pair of dual equations:
\begin{tcolorbox}
\begin{align}
\exists_M R \, \tensorialand \, \exists_N S  \quad & \viso \quad \exists_{M\tensorialand N} \,  (R\tensorialand S)  \tag{$c$}\label{eqn:c}
\\
\forall_M R \, \tensorialor \, \forall_N S  \quad & \viso \quad  \forall_{M\tensorialor N} \,  (R\tensorialor S) \tag{$d$}\label{eqn:d}
\end{align}
\end{tcolorbox}
\noindent
where $M:A\dto C$, $N:B\dto D$ and $R\refs A$, $S\refs B$ in $p$ for equation~(\ref{eqn:c})
while $M:C\dto A$, $N:D\dto B$ and $R\refs A$, $S\refs B$ in $q$ for equation~(\ref{eqn:d}).
This pair of dual formulas is fundamental:
in particular, it has the remarkable property of unifying equation~(\ref{equation/exists-vs-tensor}) with the other equation
\begin{equation}\label{equation/forall-vs-multimap}
(\, \exists_{M} R \, ) \, \multimap \, (\, \forall_{N} \, S\, ) \quad \viso \quad \forall_{M\multimap N} \, (R\multimap S)
\end{equation}
valid in~$p:\SUBSET\to\Rel$ and more generally in any bifibration $p:\Ecategory\to\Bcategory$ which is at the same time 
a symmetric monoidal closed refinement system (see Prop~2.4 in \cite{mz15isbell}).
To that purpose, one needs to replace the tensor product~$\tensorialor:q\times q\to q$
in formula~(\ref{eqn:d}) by the action of $q$ on $p$ written (on purpose) with the same notation $\tensorialor:q\times p\to p$.
Understood in this alternative way, the formula~(\ref{eqn:d}) is not equivalent anymore to equation~(\ref{equation/exists-vs-tensor})
but to equation~(\ref{equation/forall-vs-multimap}) where the refinement $R\multimap S$
and change-of-basis morphism~$M\multimap N$
$$R\multimap S \refs A\multimap B
\quad\quad
M\multimap N:C\multimap D\dto A\multimap B$$
are decomposed in the same way as we did in~(\ref{equation/chiral-decomposition})
for the implication formula, using the formalism of monoidal closed chiralities:
$$
R^{\ast}\tensorialor S\refs A^{\ast}\tensorialor B
\quad\quad
M^{\ast}\tensorialor N: C^{\ast}\tensorialor D\dto A^{\ast}\tensorialor B
$$
where $M:A\dto C$ and $N:D\dto B$ and $R\refs A$ in~$q$, 
$S\refs B$ in~$p$.
On the other hand, we have seen in~(\ref{equation/inclusion-not-equality})
that we have two canonical morphisms which are not invertible in general:
\begin{tcolorbox}
\begin{align}
\forall_M R \, \tensorialand \, \forall_N S  \quad & \to \quad\forall_{M\tensorialand N} \,  (R\tensorialand S)  \tag{$e$}\label{eqn:e}
\\
\exists_{M\tensorialor N} \,  (R\tensorialor S) \quad & \to \quad \exists_M R \, \tensorialor \, \exists_N S \tag{$f$}\label{eqn:f}
\end{align}
\end{tcolorbox}
\noindent
where $M:C\dto A$, $N:D\dto B$, $R\refs A$, $S\refs B$ in $p$ in equation~(\ref{eqn:e})
and where $M:A\dto C$, $N:B\dto D$, $R\refs A$, $S\refs B$ in $q$ in equation~(\ref{eqn:f}).
One main achievement of our approach is to recover the dualities of linear logic 
in categorical situations such as the subset hyperdoctrine on~$\Set$
or the presheaf hyperdoctrine on~$\Cat$, which are traditionally
seen as intuitionistic.
We will see in particular (\S\ref{section/comparison-with-linear-logic})
that the formulas~(\ref{eqn:a})--(\ref{eqn:f}) are bifibrational generalisations of familiar distributivity laws of linear logic.

\paragraph{The identity predicate in the subset bifibration.}
As defined by Lawvere, a hyperdoctrine is a pseudofunctor $\predicates{}:\Bcategory^\op\to\Cat$
from a cartesian closed category~$\Bcategory$ 
whose fibers $\predicates{A}$ are themselves cartesian closed categories,
and such that every substitution functor $\predicates{f}$ has a left adjoint $\Sigma_f$
and a right adjoint~$\Pi_f$.
Given such a hyperdoctrine $\predicates{}:\Bcategory^\op\to\Cat$,
Lawvere suggested to define the identity predicate $\lawvereidentity{A}\in\predicates{A\times A}$
associated to an object~$A\in\Bcategory$ as the terminal object~$\top_A\in\predicates{A}$
existentially quantified along the diagonal map $\Delta_A:A\to A\times A$, as follows:
$$
\lawvereidentity{A} \quad \defeq \quad \Sigma_{\Delta_A} \, (\, \top_A \, )
$$
In the case of the subset hyperdoctrine~$\predicates{}$, another construction 
of the identity predicate is possible, starting this time 
from the subset bifibration~$p:\SUBSET\to\Rel$ on sets and relations.
Consider the binary relation $curry(id_A):1\dto A\times A$ obtained 
by currying the identity relation~$\id_A:A\dto A$, where $1$ is the singleton set.
Then, define the identity predicate $\relationalidentity{A}$ as the singleton subset $1\in\predicates{1}$ existentially quantified along~$curry(id_A)$:
$$
\relationalidentity{A} \quad \defeq \quad \exists_{\,curry(id_A)} \, (\, 1 \, )
$$
The identity predicates $\lawvereidentity{A}$ and $\relationalidentity{A}$ coincide
in the case of the subset hyperdoctrine, but we will see (\S\ref{section/identity-type})
that they differ in the case of the presheaf hyperdoctrine, and that $\relationalidentity{A}$ appears
to be the appropriate definition in that case.

\paragraph{Plan of the paper.}
After this long and detailed introduction, we explain in \S\ref{section/presheaf-bifibration}
how to adapt smoothly all the results established here for the subset hyperdoctrine on~$\Set$
to the presheaf hyperdoctrine on~$\Cat$.
We then come back to the question of identity in \S\ref{section/identity-type}, explaining how the definition $\relationalidentity{A}$ lifts naturally to the presheaf hyperdoctrine and more generally to any monoidal closed bifibration.
In \S\ref{section/string-diagrams} we introduce a string diagram notation for presheaves (highly reminiscent of C.~S.~Peirce's ``existential graphs'' for predicate logic) which is derived from the decomposition of monoidal closed bifibrations as monoidal closed bifibration chiralities.
Finally, in \S\ref{section/comparison-with-linear-logic} we explain in what sense 
the formulas (\ref{eqn:a})--(\ref{eqn:f}) extend to bifibrations some familiar distributivity principles of linear logic.

\paragraph{Related works.}

The literature contains several different answers to the question of what exactly it means to combine ``linear logic with bifibrations''.
The approach that we develop here as well as in prior work \cite{mz15popl,mz15isbell} is to consider a functor which is both (symmetric) monoidal closed and a bifibration, with these two structures provided independently (but generating a rich interaction).
The same approach is taken in Hasegawa's work \cite{hasegawa99} on logical predicates for models of linear logic, as well as in Katsumata's work on logical predicates for computational effects \cite{katsumata2005}.
Both build on Hermida's thesis \cite{hermida93thesis} which considered a notion of ``fibred-ccc'', although a subtle difference with Hermida's work is that the latter is phrased in terms of \emph{fibred adjunctions} \cite{borceux2}, meaning that the functors associated to the cartesian closed structure are explicitly required to preserve cartesian morphisms.
That idea can also be seen as the background for Birkedal, Møgelberg, and Petersen's work on linear Abadi-Plotkin logic \cite{birkedalmogelbergpetersen-linear-abadi-plotkin}, as well as Shulman's definition \cite{shulman08} of ``monoidal bifibration'' that asks for the tensor product operation $\otimes : \Ecategory \times \Ecategory \to \Ecategory$ of the total category to preserve both cartesian and opcartesian morphisms.
Our perspective is that when such preservation properties hold, they should rather be seen as a consequence of an underlying adjunction between refinement systems \cite[Prop.~2.4]{mz15isbell}.
Most importantly, the requirement that the tensor product operation preserves cartesian morphisms is violated for the key models introduced in this paper, and in general we only have the non-invertible principle (\ref{eqn:e}).

\section{The presheaf bifibration on distributors}\label{section/presheaf-bifibration}
In this section, we explain how to adapt to the presheaf hyperdoctrine~$\predicates{}$
on~$\Cat$ everything which was established in the introduction
for the subset hyperdoctrine~$\powerset{}$ on~$\Set$.
The first step is to replace the category~$\Rel$ by the bicategory $\Dist$ (introduced by B\'enabou \cite{benabou-distributeurs,benabou-streicher}) 
whose objects are small categories $A,B$, 1-cells $M : A \dto B$ are \emph{distributors} defined as functors 
$$M \quad : \quad B^\op \times A \quad \morph{} \quad \Set$$
and 2-cells are natural transformations between distributors.
The identity 1-cell $B \dto B$ is defined as the hom functor $B^\op\times B \to \Set$, which we denote $\id_B$, 
and the composition of two distributors $M : A \dto B$ and $N : B \dto C$ is defined 
using the coend formula
$$
N\circ M \quad = \quad (c,a) \quad \mapsto \quad \int^{b\in B} \,\, N(c,b)\timesSet M(b,a)
$$
The category~$\SUBSET$ is then replaced by the category~$\PSH$
whose objects are the distributors $R : 1 \dto A$, $S : 1 \dto B$ (i.e., contravariant presheaves),
and whose morphisms $R \to S$ are pairs $(M,\alpha)$ consisting of a distributor $M : A \dto B$ 
and of a natural transformation $\alpha : M\circ R \Rightarrow S$, which may be depicted as
$$\scalebox{1}{\vcenteredhbox{\xymatrix{&1\ar[dl]_R\ar[dr]^S & \\ A \ar[rr]_M &\utwocell<\omit>{\alpha}& B}}}$$
Similarly, the category~$\coSUBSET$ is replaced by the category~$\coPSH$
whose objects are the distributors $R : A \dto 1$, $S : B \dto 1$ (i.e., covariant presheaves),
and whose morphisms $R \to S$ are pairs $(M,\alpha)$ consisting of a distributor $M : A \dto B$ 
together with a natural transformation $\alpha :  S\circ M \Rightarrow R$, which may be depicted as
$$\scalebox{1}{\vcenteredhbox{\xymatrix{&1 & \\ B\ar[ur]^{S} &\utwocell<\omit>{\alpha}& A\ar[ul]_{R}\ar[ll]^{M}}}}$$
Note that $\PSH$ and $\coPSH$ are bicategories just like $\Dist$, but we prefer to consider them
as categories for simplicity.
The two obvious functors
$$
p:\PSH\to\Dist \quad\quad q:\coPSH\to\Dist
$$
are opfibrations, and they define together a bifibration chirality,
$$
\xymatrix @-1.9pc {
\PSH\ar[ddddd]_-{p}
\ar@<.5ex>[rrrrrrrr]^-{(-)^{\ast}}
&&&&&&&&
\ar@<.5ex>[llllllll]^-{^{\ast}(-)}
\coPSH^\op\ar[ddddd]^-{q^{\,op}}
\\
\\
\\
\\
\\
\Dist
\ar@<.5ex>[rrrrrrrr]^-{(-)^{\ast}}
&&&&&&&&
\ar@<.5ex>[llllllll]^-{^{\ast}(-)}
\Dist^\op
}
$$
where the equivalence between $\PSH$ and $\coPSH^{\,op}$
transports every small category~$A$ to its opposite category~$A^*=A^\op$,
every distributor $M:A\dto B$ to the opposite distributor $M^\ast:B^\op\dto A^\op$
defined as the functor
$$
M^\ast \hspace{.5em} = \hspace{.5em} (a,b) \hspace{.5em} \mapsto \hspace{.5em} M\,(b,a) \hspace{.5em} : \hspace{.5em} 
A^{\,op} \times B \hspace{.5em} \morph{} \hspace{.5em} \Set
$$
and every contravariant presheaf~$R:1\dto A$ to the covariant presheaf $R^{\ast}:A^\op\dto 1$.
The equivalence between the refinement systems~$p$ and $q^\op$
follows from the bijective correspondence of 2-cells in~$\Dist$:
$$
\scalebox{1}{\vcenteredhbox{\xymatrix{&1\ar[dl]_R\ar[dr]^S & \\ A \ar[rr]_M &\utwocell<\omit>{}& B}}}
\quad\leftrightarrow\quad
\scalebox{1}{\vcenteredhbox{\xymatrix{&1 & \\ A^\op\ar[ur]^{R^*} &\utwocell<\omit>{}& B^\op\ar[ul]_{S^*}\ar[ll]^{M^*}}}}
$$
The induced bifibrational structure on $\PSH \to \Dist$ may be explicitly defined using coends and ends, as categorical analogues of the corresponding formulas for $\SUBSET \to \Rel$:
\begin{align}\label{equation/(co)end-formulas}
\begin{split}
\exists_M \, R \quad &=\quad b\quad \mapsto \quad \int^{a\in A} \,\, M(b,a)\timesSet R(a)
\\
\forall_M \, S  \quad &=\quad a\quad \mapsto \quad \int_{b\in B} \,\, M(b,a) \impliesSet S(b)
\end{split}
\end{align}
The following property is fundamental:
\begin{theorem}
The refinement system $p : \PSH  \morph{}  \Dist$ is symmetric monoidal closed
with tensor product and implication of the category~$\PSH$ defined as
\begin{equation}\label{equation/PSHsmc}
\!\!\!
\begin{array}{ccccccc}
R\otimes S & \!\!\! = \!\!\! & (a,b) & \! \mapsto \! & Ra \timesSet Sb & \!\!\! : \!\!\! & 1\dto A\times B
\\
R\multimap S & \!\!\! = \!\!\!  & (a,b) &\! \mapsto \! & Ra\impliesSet Sb & \!\!\! : \!\!\!  & 1\dto A^\op\times B
\end{array}
\end{equation}
where $R:1\dto A$ and $S:1\dto B$ are contravariant presheaves.
\end{theorem}
\noindent
Note that the implication $R\multimap S$ in $\PSH$
is transported by $p$ to the implication $A\multimap B$ defined as $A^\op\times B$
in the compact closed bicategory~$\Dist$.
From this follows that the category $\coPSH$ together with the category $\PSH$ defines
a symmetric monoidal closed chirality
$$
\Acategory=(\PSH,\tensorialand,\tensorialtrue)
\quad\quad
\Bcategory=(\coPSH,\tensorialor,\tensorialfalse)
$$
which thus satisfies formulas~(\ref{eqn:a})--(\ref{eqn:f}) stated in the introduction.
Now, let us recall that in Lawvere's presheaf hyperdoctrine 
$$\predicates{A} = [A^\op,\Set]$$
the substitution operation
$$
\substitution{F} \quad : \quad [B^\op,\Set] \quad \morph{} \quad  [A^\op,\Set]
$$
along a functor $F : A \to B$ is defined by precomposition 
$$
\substitution{F} S\quad =\quad a \quad \mapsto\quad S(F a)
$$
while the quantifiers 
$$
\Sigma_F,\Pi_F \quad : \quad [A^\op,\Set] \quad \morph{} \quad  [B^\op,\Set]
$$
may be defined as coends/ends:
\begin{align*}
\begin{split}
\Sigma_F \, R \quad &=\quad b\quad \mapsto \quad \int^{a\in A} \,\, \hom_B(b,Fa)\timesSet R(a)
\\
\Pi_F \, R  \quad &=\quad b\quad \mapsto \quad \int_{a\in A} \,\, \hom_B(Fa,b) \impliesSet R(a)
\end{split}
\end{align*}
In essentially the same way as we saw earlier for the subset hyperdoctrine over sets, the presheaf hyperdoctrine can be decomposed into a pair of bifibrations
$$
\embP{p}:\embP{\Psh}\to\Cat
\quad\quad\quad
\embN{p}:\embN{\Psh}\to\Cat^\op
$$
where:
\begin{itemize}
\item $\embP{\Psh}$ has as objects pairs $(A,R)$ consisting of a category $A$ together with contravariant presheaf $R : A^\op \to \Set$, and morphisms $(F,\alpha) : (A,R) \to (B,S)$ consisting of a pair of a functor $F : A \to B$ together with a natural transformation $\alpha : R \Rightarrow S \circ F^\op$;
\item $\embN{\Psh}$ has as objects pairs $(A,R)$ consisting of a category $A$ together with a covariant presheaf $R : A \to \Set$, and morphisms $(F,\alpha) : (A,R) \to (B,S)$ consisting of a pair of a functor $F : B \to A$ together with a natural transformation $R \circ F \Rightarrow S$;
\item $\embP{p}$ and $\embN{p}$ are the evident projections.
\end{itemize}
Moreover, there are a pair of embedding functors
$$\embP{\emb}:\Cat \to \Dist
\quad\quad\quad
\embN{\emb}:\Cat^\op \to \Dist$$
acting as the identity on objects and sending a functor $F : A \to B$ to the respective distributors
$$
\begin{array}{ccl}
\embP{F} & = & (b,a) \quad \mapsto \quad \hom_B(b,Fa) \quad:\quad A \dto B
\\
\embN{F} & = & (a,b) \quad \mapsto \quad \hom_B(Fa,b) \quad:\quad B \dto A
\end{array}
$$
with the property that 
\begin{theorem}\label{thm/reconstruction-presheaves}
The bifibrations $\embP{p} : \embP{\Psh}\to\Cat$ and $\embN{p} : \embN{\Psh}\to\Cat^\op$ are pullbacks of $p : \PSH \to \Dist$ along the functors $\embP{\emb}$ and $\embN{\emb}$, respectively.
\end{theorem}
\noindent
\noindent
Once again, this theorem implies that the triple adjunction
$$
\Sigma_F \quad \dashv \quad \substitution{F} \quad \dashv \quad \Pi_F
$$
of the presheaf hyperdoctrine may be reduced to a pair of adjunctions
$$
\exists_{\embP{F}} \quad \dashv \quad \forall_{\embP{F}}  \quad = \quad \exists_{\embN{F}} \quad \dashv \quad\forall_{\embN{F}}
$$
of the bifibration $p : \PSH \to \Dist$.

\medbreak

The other important logical ingredient in Lawvere's original definition of a hyperdoctrine is the cartesian closed structure of each category of predicates $\predicates{A}$.
Here we note that the well-known ccc structure on presheaf categories may be further decomposed using the monoidal closed structure of the presheaf bifibration on distributors, beginning with the following elementary observation (recalled from \cite{mz15popl,mz15isbell}):
\begin{proposition}\label{prop:dayconstruction}
If $\Ecategory \to \Bcategory$ is a monoidal closed refinement system which is also a bifibration, then every monoid
$$(A, m : A \otimes A \to A, e : 1 \to A) \quad\in\quad \Bcategory$$
in the basis determines a monoidal closed structure on the fiber $\Ecategory_A$, where the tensor and implication are defined for all $R,S \refs A$ by
\begin{align*}
R \otimes_A S &\defeq \pushverbatim{m}(R\otimes S)\\
R \multimap_A S &\defeq \pullverbatim{curry(m)}(R \multimap S)
\end{align*}
and the tensor unit is defined by $1_A \defeq \pushverbatim{e}1$.
\end{proposition}
\noindent
Every category is a comonoid $(A,\Delta_A : A \to A \times A, !_A : A \to 1)$ in $\Cat$, and is hence transported by the functor $\embN{\emb}$ to a monoid:
$$(A, \embN{\Delta}_A : A\times A \dto A, \embN{!}_A : 1 \dto A) \quad\in\quad \Dist$$
The fiber of $A$ in $p : \PSH \to \Dist$ is thus automatically endowed with a monoidal closed structure by \Cref{prop:dayconstruction},
$$
R\wedge S \defeq \pushD{\embN{\Delta}_A}(R \otimes S) = \pullD{\embP{\Delta}_A}(R \otimes S)
\qquad
\top \defeq \pushD{\embN{!}_A}1 = \pullD{\embP{!}_A}1
$$
$$
R\supset S \defeq \pullD{curry(\embN{\Delta}_A)}(R \multimap S)
$$
and it is straightforward to verify using equations (\ref{equation/(co)end-formulas}) and (\ref{equation/PSHsmc}) that this monoidal closed structure $(\wedge_A,\top_A,\supset_A)$ is isomorphic to the usual ccc structure on the presheaf category $[A^\op,\Set]$.

\section{The problem of identity}
\label{section/identity-type}

We now turn to Lawvere's abstract definition of the identity predicate $\lawvereidentity{A} \defeq \Sigma_{\Delta_A} \, (\, \top_A \, )$ in an arbitrary hyperdoctrine \cite{lawvere70}.
In the presheaf hyperdoctrine this definition yields
$$
\lawvereidentity{A} \,\, = \,\, (a_1,a_2)\,\,\mapsto\,\, \int^{a\in A} \hom_A(a_1,a) \timesSet \hom_A(a_2,a)
$$
which does not seem to give an appropriate notion of identity (any pair of objects are ``equated'' so long as they can be completed to a cospan), even in the case when the category $A$ is a groupoid.
Lawvere remarked that a more natural choice of generalized ``identity predicate'' on a category $A$ within the presheaf hyperdoctrine is the functor $\hom_A : A^\op \times A \to \Set$.
Our first observation is that this version of the identity predicate may be easily defined as a pushforward in the symmetric monoidal closed refinement system $p : \PSH \to \Dist$ by 
$$\relationalidentity{A}= \graphD{\id_A} \defeq \pushD{curry(\id_A)}\, (\, 1 \, )$$
and that more generally we can recover the presheaf associated to a distributor $M : A \dto B$ by the formula
\begin{align*}
\graphD{M} &\refs A^\op \times B \\
\graphD{M} &\defeq \pushD{curry(M)}\, (\, 1 \, )
\end{align*}
Indeed, this abstract recipe allows us to define the ``graph'' of a morphism $f : A \to B$ in a monoidal closed category $\Bcategory$ with respect to any monoidal closed bifibration $p:\Ecategory\to\Bcategory$
\begin{align*}
\graphD{f} &\refs A \multimap B \\
\graphD{f} &\defeq \pushverbatim{curry(f)}\,I
\end{align*}
where $I$ is the monoidal unit of~$\Ecategory$.
We then have
\begin{theorem}
\label{thm:pushpull-canonical}
Let $p: \Ecategory \to \Bcategory$ be a monoidal closed refinement system which is also a bifibration, and suppose given refinements $R \refs A$ and $S \refs B$ in $\Ecategory$ and a morphism $f : A \to B$ in $\Bcategory$.
Then there are isomorphisms
\begin{align}
\pushverbatim{f}R &\viso \pushverbatim{eval}(R \otimes \graphD{f}) \label{equation/push-canonical} \\
\pullverbatim{f}S &\viso \pullverbatim{dni}(S \ImpR \graphD{f}) \label{equation/pull-canonical}
\end{align}
where $eval : A \otimes (A \multimap B) \morph{} B$ is the left evaluation map, and where $dni : A \morph{} B \ImpR (A \multimap B)$ is the right currying of $eval$.
\end{theorem}
\noindent
\emph{Proof.}
Both formulas follow easily from distributivity properties discussed in the introduction together with the axioms of monoidal closed categories:
\begin{align*}
\pushverbatim{eval}(R \otimes \graphD{f}) &\viso \pushverbatim{eval}\pushverbatim{(\id_A \otimes curry(f))}(R \otimes I) \\
 &\viso \pushverbatim{eval\circ (\id_A \otimes curry(f))}R \\
 &\viso \pushverbatim{f}R \\
\pullverbatim{dni}(S \ImpR \graphD{f}) &\viso \pullverbatim{dni}\pullverbatim{\id_B \ImpR curry(f)}(S \ImpR I) \\
 &\viso \pullverbatim{dni;(\id_B \ImpR curry(f))}S \\
 &\viso \pullverbatim{f}S
\end{align*}
\noindent
Equation (\ref{equation/push-canonical}) may be compared with Lawvere's equation \cite[p.8]{lawvere70}
$$\Sigma_f R \viso \Sigma_{\pi_2} (\substitution{\pi_1}R \wedge \lawvereidentity{f})$$
where $\pi_1 : A \times B \to A$ and $\pi_2 : A \times B \to B$ are the projection maps and where the graph $\lawvereidentity{f} \in \predicates{A\times B}$ is defined by substitution along $f \times \id_B$ 
into $\lawvereidentity{B}$.
Lawvere proved that this equation holds for any hyperdoctrine satisfying Frobenius Reciprocity and a Beck-Chevalley condition, but he also explicitly observed that those conditions are violated by the presheaf hyperdoctrine.

On the other hand, equation (\ref{equation/pull-canonical}) may be thought of as an abstract generalization of Yoneda's lemma.
Indeed, one can consider the formula in the bifibration $\embP{p} : \embP{\Psh} \to \Cat$, which is also a cartesian closed refinement system \cite{mz15popl}.
Taking $f = a : 1 \to A$ to be an object of the category $A$, the two sides of (\ref{equation/pull-canonical}) expand to
$$
S(a) = \int_{a' \in A} \hom_A(a',a) \impliesSet S(a')
$$
which is the precise statement of the Yoneda lemma.

\section{A Peircean notation for presheaves as generalized predicates}
\label{section/string-diagrams}

The prolific American logician Charles Sanders Peirce developed during the late 19th and early 20th centuries a system for representing logical deductions as certain topological surgeries on diagrams he called ``existential graphs''.\footnote{To witness existential graphs in Peirce's own words and drawings, see for example his ``Prolegomena to an Apology for Pragmaticism'', published in {\em The Monist}, Vol.~16, No.~4 (October 1906), pp.~492--546, and freely available at \url{http://www.jstor.org/stable/27899680}.}
A key element of Peirce's graphical calculus was the idea of identifying the subject of two predicates by joining them with an arc.
For example, the diagram
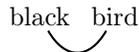
\begin{center}
\scalebox{1}{\begin{tikzpicture}
	\begin{pgfonlayer}{nodelayer}
		\node [style=rel] (0) at (-1, 0) {black};
		\node [style=rel] (1) at (0, 0) {bird};
	\end{pgfonlayer}
	\begin{pgfonlayer}{edgelayer}
		\draw [style=wire,bend right=60, looseness=1.5] (0) to (1);
	\end{pgfonlayer}
\end{tikzpicture}}
\end{center}
expresses that there is something which is both black and a bird (such as a crow).
A second key element of existential graphs was the use of an enclosing curve (which Peirce called a ``cut'' or ``sep'') to negate a proposition.
Thus the diagram
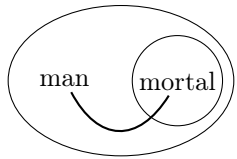
\begin{center}
\scalebox{1}{\begin{tikzpicture}
	\begin{pgfonlayer}{nodelayer}
		\node [style=rel] (0) at (-1.5, 0) {man};
		\node [style=rel] (1) at (0, 0) {mortal};
	\end{pgfonlayer}
	\begin{pgfonlayer}{edgelayer}
		\draw [style=wire,bend right=60, looseness=1.5] (0) to (1);
                \draw [style=sep] (1) circle (0.6);
                \draw [style=sep] (-0.75,0) ellipse (1.5 and 1);
	\end{pgfonlayer}
\end{tikzpicture}}
\end{center}
expresses that there does not exist a man who is not mortal, i.e., that every man is mortal.
Similarly, the diagram
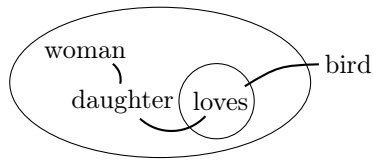
\begin{center}
\scalebox{1}{\begin{tikzpicture}
	\begin{pgfonlayer}{nodelayer}
		\node [style=rel] (0) at (-2, 0.4) {woman};
		\node [style=rel] (3) at (-1.5, -0.25) {daughter};
		\node [style=rel] (1) at (1.5, 0.25) {bird};
		\node [style=rel] (2) at (-0.2, -0.25) {loves};
	\end{pgfonlayer}
	\begin{pgfonlayer}{edgelayer}
		\draw [style=wire,bend left=30] (0) to (3);
		\draw [style=wire,bend right=45] (3) to (2);
		\draw [style=wire,bend left=15, looseness=1.25] (2) to (1);
                \draw [style=sep] (-0.25,-0.25) ellipse (0.5 and 0.5);
                \draw [style=sep] (-1,0) ellipse (2 and 1);
	\end{pgfonlayer}
\end{tikzpicture}}
\end{center}
expresses that there is some particular bird that every woman's daughter loves (the most popular bird at the park).

Geraldine Brady and Todd Trimble gave a categorical interpretation of existential graphs \cite{bradytrimble} within Joyal and Street's framework of \emph{string diagrams} for monoidal categories \cite{joyal-street-i}.
Their analysis began with a categorical axiomatization of classical first-order logic in terms of \emph{Boolean hyperdoctrines}, that is, hyperdoctrines with Boolean algebra fibers and satisfying the Beck-Chevalley condition.
They then went on to describe how to interpret the predicates of any such hyperdoctrine as 1-cells in an appropriate compact closed bicategory of Boolean-valued relations.

In this part of the paper we reveal a surprising development of Brady and Trimble's work, by explaining how the logical structure of the refinement system $p : \PSH \to \Dist$ in combination with its chiral opposite $q : \coPSH \to \Dist$ leads in a relatively straightforward way to a string diagram calculus for presheaves that is remarkably reminiscent of Peirce's existential graphs -- this despite the fact that our ``predicates'' and ``relations'' are by no means Boolean-valued!

We will assume that the reader already has some familiarity with string diagrams for monoidal categories in general, and with the standard conventions for compact closed categories (otherwise, the reader is encouraged to read \cite{selinger-survey} for a quick introduction).
For example, following those conventions, the diagram
$$\vcenteredhbox
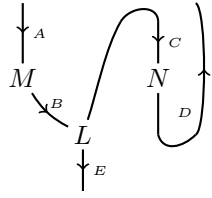
{\begin{tikzpicture}
	\begin{pgfonlayer}{nodelayer}
		\node [style=rel] (0) at (0.5, 0.75) {$N$};
		\node [style=rel] (1) at (-1.3, 0.75) {$M$};
		\node [style=rel] (2) at (-0.5, 0) {$L$};
		\node [style=none] (3) at (-1.3, 1.75) {};
		\node [style=none] (4) at (0, 1.5) {};
		\node [style=none] (5) at (-0.5, -0.75) {};
		\node [style=none] (6) at (0.5, 0) {};
		\node [style=none] (7) at (0.5, 1.5) {};
		\node [style=none] (8) at (1, 0) {};
		\node [style=none] (9) at (1, 1.75) {};
	\end{pgfonlayer}
	\begin{pgfonlayer}{edgelayer}
		\draw [style=directed] (3.center) to node [right] {\tiny$A$} (1);
		\draw [style=directed, bend right=15] (1) to node [near start,right] {\tiny$B$} (2);
		\draw [style=directed] (2) to node [right] {\tiny$E$} (5.center);
		\draw [style=directed] (7.center) to node [right] {\tiny$C$} (0);
		\draw [style=wire] (0) to (6.center);
		\draw [style=wire, in=90, out=60, looseness=1.25] (4.center) to (7.center);
		\draw [style=wire, in=-120, out=72, looseness=0.75] (2) to (4.center);
		\draw [style=wire, in=-135, out=-75, looseness=1.25] (6.center) to (8.center);
		\draw [style=directed, bend right=60, looseness=0.25] (8.center) to node [near start,left] {\tiny$D$} (9.center);
	\end{pgfonlayer}
\end{tikzpicture}}$$
represents the distributor $L \circ (M \otimes N^*) : A \times D^\op \dto E$ obtained by composing distributors $M : A \dto B$, $N : C \dto D$, and $L : B \times C^\op \dto E$ in the indicated way.
Observe that here we read the diagram from top-to-bottom and left-to-right, while we place labels indicating the underlying categories to the left-hand side of each oriented wire.
More abstractly, in topological terms these conventions can be said to rely on the assumption that the surface in which the diagram of the distributor is embedded (in this case, the page) is equipped with an orientation.

To represent a contravariant presheaf $R : A^\op \to \Set$ seen as a refinement $R \refs A$ in $p : \PSH \to \Dist$, we lay it out just as one would an ordinary distributor $R : 1 \dto A$, but framed by a light blue background to indicate that we view it as an object of $\PSH$.
Therefore, the operations of taking the tensor product of presheaves or the pushforward along a distributor,
$$
\infer{R \otimes S \refs A \times B}{R \refs A & S \refs B}
\qquad
\infer{\pushD{M}R \refs B}{R \refs A & M : A \dto B}
$$
which can be defined respectively in terms of horizontal and vertical composition of 1-cells in $\Dist$, are displayed like so:
$$
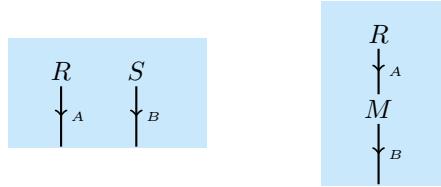
\vcenteredhbox{\begin{tikzpicture}
	\begin{pgfonlayer}{nodelayer}
		\node [style=rel] (0) at (1, 1) {$S$};
		\node [style=none] (1) at (1, 0) {};
		\node [style=rel] (3) at (0, 1) {$R$};
		\node [style=none] (4) at (0, 0) {};
	\end{pgfonlayer}
	\begin{pgfonlayer}{edgelayer}
		\draw [style=directed] (0) to node [right] {\tiny$B$} (1.center);
		\draw [style=directed] (3) to node [right] {\tiny$A$} (4.center);
	\end{pgfonlayer}
	\begin{pgfonlayer}{framelayer}
		\path[fill=bleu] 
		($(current bounding box.south west) - (0.5,0)$) rectangle 
		($(current bounding box.north east) + (0.5,0.25)$); 
	\end{pgfonlayer}
\end{tikzpicture}}
\qquad\qquad
\vcenteredhbox{\begin{tikzpicture}
	\begin{pgfonlayer}{nodelayer}
		\node [style=rel] (3) at (0, 1) {$R$};
		\node [style=rel] (0) at (0, 0) {$M$};
		\node [style=none] (1) at (0, -1) {};
	\end{pgfonlayer}
	\begin{pgfonlayer}{edgelayer}
		\draw [style=directed] (3) to node [right] {\tiny$A$} (0);
		\draw [style=directed] (0) to node [right] {\tiny$B$} (1.center);
	\end{pgfonlayer}
	\begin{pgfonlayer}{framelayer}
		\path[fill=bleu] 
		($(current bounding box.south west) - (0.5,0)$) rectangle 
		($(current bounding box.north east) + (0.5,0.25)$); 
	\end{pgfonlayer}
\end{tikzpicture}}
$$
Let us note here that there is a topological interpretation of the refinement relation, in the sense that \emph{a diagram embedded in a surface refines its boundary.}
As for implication between presheaves or the pullback along a distributor,
$$
\infer{R \multimap S \refs A^\op \times B}{R \refs A & S \refs B}
\qquad
\infer{\pullD{M}S \refs A}{M : A \dto B & S \refs B}
$$
we base our conventions on the equations
$$
R\multimap S \viso {}^*(S^* \tensorialand R)
\qquad
\pullD{M}S \viso {}^*(\pushD{M^*}S^*)
$$
and draw the following diagrams:
$$
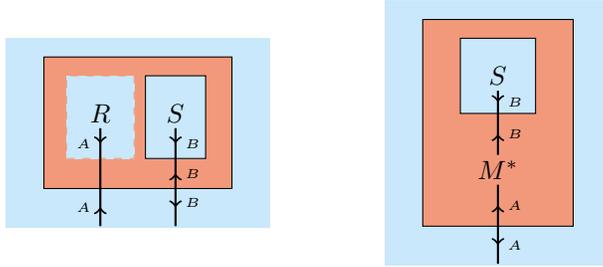
\vcenteredhbox{\begin{tikzpicture}
	\begin{pgfonlayer}{nodelayer}
		\node [style=rel] (0) at (0.75, 1.75) {$S$};
		\node [style=none] (1) at (0.75, 1.15) {};
		\node [style=none] (2) at (0.75, 0.75) {};
		\node [style=rel] (3) at (-0.25, 1.75) {$R$};
		\node [style=none] (7) at (-0.25, 1.15) {};
		\node [style=none] (4) at (-0.25, 0.75) {};
		\node [style=none] (5) at (-0.25, 0.25) {};
		\node [style=none] (6) at (0.75, 0.25) {};
	\end{pgfonlayer}
	\begin{pgfonlayer}{edgelayer}
		\draw [style=directed] (0) to node [right] {\tiny$B$} (1.center);
		\draw [style=directed] (2.center) to node [right] {\tiny$B$} (1.center);
		\draw [style=directed] (5.center) to node [left] {\tiny$A$} (4.center);
		\draw [style=directed] (2.center) node [below right] {\tiny$B$} to (6.center);
		\draw [style=directed] (3) to node [left] {\tiny$A$} (7.center);
		\draw [style=wire] (7.center) to (4.center);
	\end{pgfonlayer}
	\begin{pgfonlayer}{boxlayer}
		\draw [style=sep,fill=rouge] (-1, 0.75) rectangle (1.5, 2.5);
		\draw [style=ghost,fill=bleu] (-0.7, 1.15) rectangle (0.2, 2.25);
		\draw [style=sep,fill=bleu] (0.35, 1.15) rectangle (1.15,2.25);
	\end{pgfonlayer}
	\begin{pgfonlayer}{framelayer}
		\path[fill=bleu] 
		($(current bounding box.south west) - (0.5,0)$) rectangle 
		($(current bounding box.north east) + (0.5,0.25)$); 
	\end{pgfonlayer}
\end{tikzpicture}}
\qquad\qquad
\vcenteredhbox{\begin{tikzpicture}
	\begin{pgfonlayer}{nodelayer}
		\node [style=rel] (0) at (-1.25, 0) {$M^*$};
		\node [style=none] (2) at (-1.25, 0.75) {};
		\node [style=rel] (7) at (-1.25, 1.25) {$S$};
		\node [style=none] (8) at (-1.25, -0.75) {};
		\node [style=none] (9) at (-1.25, -1.25) {};
	\end{pgfonlayer}
	\begin{pgfonlayer}{edgelayer}
		\draw [style=directed] (7) to node [right] {\tiny$B$} (2.center);
		\draw [style=directed] (8.center) to node [right] {\tiny$A$} (0);
		\draw [style=directed] (0) to node [right] {\tiny$B$} (2.center);
		\draw [style=directed] (8.center) to node [right] {\tiny$A$} (9.center);
	\end{pgfonlayer}
	\begin{pgfonlayer}{boxlayer}
		\draw [style=sep,fill=rouge] (-2.25, -0.75) rectangle (-0.25,2);
		\draw [style=sep,fill=bleu] (-1.75, 0.75) rectangle (-0.75, 1.75);
	\end{pgfonlayer}
	\begin{pgfonlayer}{framelayer}
		\path[fill=bleu] 
		($(current bounding box.south west) - (0.5,0)$) rectangle 
		($(current bounding box.north east) + (0.5,0.25)$); 
	\end{pgfonlayer}
\end{tikzpicture}}
$$
Here the dualization operation $(-)^* : \PSH \to \coPSH^{\op(0,1)}$ and its inverse ${}^*(-)  : \coPSH^{\op(0,1)} \to \PSH$ are being represented as ``functorial boxes'' \cite{mellies-fboxes}, which take the mirror image of the boundary wires across the box from a blue region to a red region and vice versa, while the action
$
\owedge : \coPSH \times \PSH \to \coPSH
$
is represented by gluing a blue diagram inside a dark red region.
One subtlety is that since the dualization operations reverse the orientation of the tensor product, we must therefore read horizontal juxtaposition in the red region (corresponding to the action $\owedge$) running right-to-left rather than left-to-right.
Also, it is worth noting that these conventions ensure that inside any red region there is always exactly one boxed (i.e., negated) blue region, which can be seen as a sort of intuitionistic restriction on Peirce's system.

We next deduce some equations on the diagrams that are implied by the axioms of a monoidal closed bifibration chirality.
Equations between purely positive formulas such as
$$\pushD{N}\pushD{M}R \viso \pushD{N \circ M}R \quad\text{and}\quad\pushD{M\otimes N}(R\otimes S) \viso \pushD{M}R \otimes \pushD{N}S$$
are geometrically manifest using these conventions, just as the axioms of monoidal categories are geometrically manifest using ordinary string diagrams.
On the other hand, since dualization is an involutive operation, it is also always possible to add or remove an annulus around the diagram of a contravariant or covariant presheaf without changing its meaning:
\begin{align}
\label{equation/doublesep}
\begin{split}
\vcenteredhbox{\scalebox{0.8}{\begin{tikzpicture}
	\begin{pgfonlayer}{nodelayer}
		\node [style=rel] (0) at (0.75, 1.75) {$R$};
		\node [style=none] (2) at (0.75, 0.75) {};
	\end{pgfonlayer}
	\begin{pgfonlayer}{edgelayer}
		\draw [style=directed] (0) to (2.center);
	\end{pgfonlayer}
	\begin{pgfonlayer}{framelayer}
		\path[fill=bleu] 
		($(current bounding box.south west) - (0.5,0)$) rectangle 
		($(current bounding box.north east) + (0.5,0.25)$); 
	\end{pgfonlayer}
\end{tikzpicture}}}
 \quad&=\quad
\vcenteredhbox{\scalebox{0.8}{\begin{tikzpicture}
	\begin{pgfonlayer}{nodelayer}
		\node [style=rel] (0) at (0.75, 1.75) {$R$};
		\node [style=none] (1) at (0.75, 1.25) {};
		\node [style=none] (2) at (0.75, 0.75) {};
		\node [style=none] (6) at (0.75, 0.5) {};
	\end{pgfonlayer}
	\begin{pgfonlayer}{edgelayer}
		\draw [style=wire] (0) to (1.center);
		\draw [style=wire] (2.center) to (1.center);
		\draw [style=directed] (2.center) to (6.center);
	\end{pgfonlayer}
	\begin{pgfonlayer}{boxlayer}
		\draw [style=sep,fill=rouge] (0, 1) rectangle (1.5, 2.5);
		\draw [style=sep,fill=bleu] (0.25, 1.25) rectangle (1.25,2.25);
	\end{pgfonlayer}
	\begin{pgfonlayer}{framelayer}
		\path[fill=bleu] 
		($(current bounding box.south west) - (0.5,0)$) rectangle 
		($(current bounding box.north east) + (0.5,0.25)$); 
	\end{pgfonlayer}
\end{tikzpicture}}} \\
\vcenteredhbox{\scalebox{0.8}{\begin{tikzpicture}
	\begin{pgfonlayer}{nodelayer}
		\node [style=rel] (0) at (0.75, 1.75) {$S$};
		\node [style=none] (1) at (0.75, 1.25) {};
		\node [style=none] (2) at (0.75, 0.75) {};
		\node [style=none] (6) at (0.75, 0.5) {};
	\end{pgfonlayer}
	\begin{pgfonlayer}{edgelayer}
		\draw [style=wire] (1.center) to (0);
		\draw [style=wire] (1.center) to (2.center);
		\draw [style=directed] (6.center) to (1.center);
	\end{pgfonlayer}
	\begin{pgfonlayer}{boxlayer}
		\draw [style=sep,fill=bleu] (0, 1) rectangle (1.5, 2.5);
		\draw [style=sep,fill=rouge] (0.25, 1.25) rectangle (1.25,2.25);
	\end{pgfonlayer}
	\begin{pgfonlayer}{framelayer}
		\path[fill=rouge] 
		($(current bounding box.south west) - (0.5,0)$) rectangle 
		($(current bounding box.north east) + (0.5,0.25)$); 
	\end{pgfonlayer}
\end{tikzpicture}}}
 \quad&=\quad
\vcenteredhbox{\scalebox{0.8}{\begin{tikzpicture}
	\begin{pgfonlayer}{nodelayer}
		\node [style=rel] (0) at (0.75, 1.75) {$S$};
		\node [style=none] (2) at (0.75, 0.75) {};
	\end{pgfonlayer}
	\begin{pgfonlayer}{edgelayer}
		\draw [style=directed] (2.center) to (0);
	\end{pgfonlayer}
	\begin{pgfonlayer}{framelayer}
		\path[fill=rouge] 
		($(current bounding box.south west) - (0.5,0)$) rectangle 
		($(current bounding box.north east) + (0.5,0.25)$); 
	\end{pgfonlayer}
\end{tikzpicture}}}
\end{split}
\end{align}
Seen in this way, the important distributivity law
$$\pushD{M}R \multimap \pullD{N}S \viso \pullD{M\multimap N}(R\multimap S)$$
simply removes an annulus, pushes one component ($M$) outside the blue region, and places the annulus back in another location:
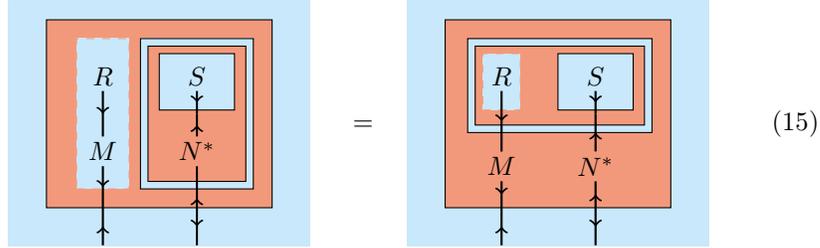
\begin{equation}\label{equation/forall-vs-multimap/stringy}
\vcenteredhbox{\scalebox{1}{\begin{tikzpicture}
	\begin{pgfonlayer}{nodelayer}
		\node [style=rel] (0) at (1, 2.5) {$S$};
		\node [style=none] (1) at (1, 2) {};
		\node [style=rel] (8) at (1, 1.5) {$N^*$};
		\node [style=none] (2) at (1, 0.75) {};
		\node [style=none] (9) at (1, 1) {};
		\node [style=none] (10) at (1, 1.25) {};
		\node [style=rel] (3) at (-0.25, 2.5) {$R$};
		\node [style=rel] (7) at (-0.25, 1.5) {$M$};
		\node [style=none] (4) at (-0.25, 0.75) {};
		\node [style=none] (5) at (-0.25, 0.25) {};
		\node [style=none] (6) at (1, 0.25) {};
	\end{pgfonlayer}
	\begin{pgfonlayer}{edgelayer}
		\draw [style=directed] (0) to (1.center);
		\draw [style=directed] (2.center) to (9.center);
		\draw [style=wire] (9.center) to (10.center);
		\draw [style=wire] (10.center) to (8);
		\draw [style=directed] (8) to (1.center);
		\draw [style=directed] (5.center) to (4.center);
		\draw [style=directed] (2.center) to (6.center);
		\draw [style=directed] (3) to (7);
		\draw [style=directed] (7) to (4.center);
	\end{pgfonlayer}
	\begin{pgfonlayer}{boxlayer}
		\draw [style=sep,fill=rouge] (-1, 0.75) rectangle (2, 3.25);
		\draw [style=sep,fill=bleu] (0.25, 1) rectangle (1.75,3);
		\draw [style=sep,fill=rouge] (0.35, 1.1) rectangle (1.65,2.9);
		\draw [style=ghost,fill=bleu] (-0.6, 1) rectangle (0.1,3);
		\draw [style=sep,fill=bleu] (0.5, 2.05) rectangle (1.5,2.8);
	\end{pgfonlayer}
	\begin{pgfonlayer}{framelayer}
		\path[fill=bleu] 
		($(current bounding box.south west) - (0.5,0)$) rectangle 
		($(current bounding box.north east) + (0.5,0.25)$); 
	\end{pgfonlayer}
\end{tikzpicture}}}
\quad=\quad
\vcenteredhbox{\scalebox{1}{\begin{tikzpicture}
	\begin{pgfonlayer}{nodelayer}
		\node [style=rel] (0) at (1, 2.5) {$S$};
		\node [style=none] (1) at (1, 2) {};
		\node [style=rel] (8) at (1, 1.3) {$N^*$};
		\node [style=none] (2) at (1, 0.75) {};
		\node [style=none] (9) at (1, 1) {};
		\node [style=none] (10) at (1, 1.25) {};
		\node [style=rel] (3) at (-0.25, 2.5) {$R$};
		\node [style=rel] (7) at (-0.25, 1.3) {$M$};
		\node [style=none] (4) at (-0.25, 0.75) {};
		\node [style=none] (5) at (-0.25, 0.25) {};
		\node [style=none] (6) at (1, 0.25) {};
	\end{pgfonlayer}
	\begin{pgfonlayer}{edgelayer}
		\draw [style=directed] (0) to (1.center);
		\draw [style=directed] (2.center) to (8);
		\draw [style=directed] (8) to (1.center);
		\draw [style=directed] (5.center) to (4.center);
		\draw [style=directed] (2.center) to (6.center);
		\draw [style=directed] (3) to (7);
		\draw [style=directed] (7) to (4.center);
	\end{pgfonlayer}
	\begin{pgfonlayer}{boxlayer}
		\draw [style=sep,fill=rouge] (-1, 0.75) rectangle (2, 3.25);
		\draw [style=sep,fill=bleu] (-0.7, 1.75) rectangle (1.75,3);
		\draw [style=sep,fill=rouge] (-0.6, 1.85) rectangle (1.65,2.9);
		\draw [style=ghost,fill=bleu] (-0.5, 2.05) rectangle (0,2.8);
		\draw [style=sep,fill=bleu] (0.5, 2.05) rectangle (1.5,2.8);
	\end{pgfonlayer}
	\begin{pgfonlayer}{framelayer}
		\path[fill=bleu] 
		($(current bounding box.south west) - (0.5,0)$) rectangle 
		($(current bounding box.north east) + (0.5,0.25)$); 
	\end{pgfonlayer}
\end{tikzpicture}}}
\end{equation}
Finally, the two formulas (\ref{equation/push-canonical}) and (\ref{equation/pull-canonical}) derived for the identity predicates defined in \S\ref{section/identity-type} have the following geometric interpretation:
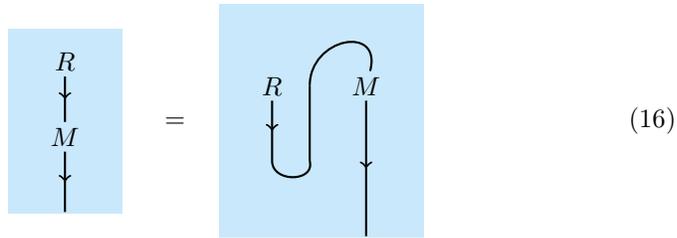
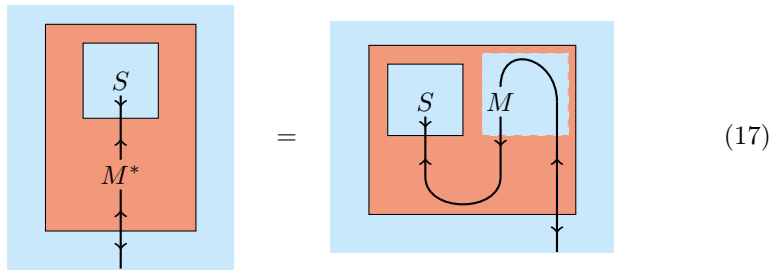
\begin{align}\label{equation/push-canonical/stringy}
\vcenteredhbox{\begin{tikzpicture}
	\begin{pgfonlayer}{nodelayer}
		\node [style=rel] (3) at (0, 1) {$R$};
		\node [style=rel] (0) at (0, 0) {$M$};
		\node [style=none] (1) at (0, -1) {};
	\end{pgfonlayer}
	\begin{pgfonlayer}{edgelayer}
		\draw [style=directed] (3) to (0);
		\draw [style=directed] (0) to (1.center);
	\end{pgfonlayer}
	\begin{pgfonlayer}{framelayer}
		\path[fill=bleu] 
		($(current bounding box.south west) - (0.5,0)$) rectangle 
		($(current bounding box.north east) + (0.5,0.25)$); 
	\end{pgfonlayer}
\end{tikzpicture}}
  &\quad=\quad \vcenteredhbox{\begin{tikzpicture}
	\begin{pgfonlayer}{nodelayer}
		\node [style=rel] (4) at (-1, 1) {$M$};
		\node [style=none] (5) at (-1, -1) {};
		\node [style=rel] (6) at (-2.25, 1) {$R$};
		\node [style=none] (7) at (-1.75, 0) {};
		\node [style=none] (8) at (-1.75, 1) {};
		\node [style=none] (9) at (-2.25, 0) {};
	\end{pgfonlayer}
	\begin{pgfonlayer}{edgelayer}
		\draw [style=directed] (4) to (5.center);
		\draw [style=wire] (7.center) to (8.center);
		\draw [style=wire,in=75, out=90, looseness=2.00] (8.center) to (4);
		\draw [style=directed] (6) to (9.center);
		\draw [style=wire,in=-75, out=-90, looseness=1.50] (9.center) to (7.center);
	\end{pgfonlayer}
	\begin{pgfonlayer}{framelayer}
		\path[fill=bleu] 
		($(current bounding box.south west) - (0.5,0)$) rectangle 
		($(current bounding box.north east) + (0.5,0.25)$); 
	\end{pgfonlayer}
\end{tikzpicture}}
\end{align}
\begin{align}\label{equation/pull-canonical/stringy}
\vcenteredhbox{\begin{tikzpicture}
	\begin{pgfonlayer}{nodelayer}
		\node [style=rel] (0) at (-1.25, 0) {$M^*$};
		\node [style=none] (2) at (-1.25, 0.75) {};
		\node [style=rel] (7) at (-1.25, 1.25) {$S$};
		\node [style=none] (8) at (-1.25, -0.75) {};
		\node [style=none] (9) at (-1.25, -1.25) {};
	\end{pgfonlayer}
	\begin{pgfonlayer}{edgelayer}
		\draw [style=directed] (7) to (2.center);
		\draw [style=directed] (8.center) to (0);
		\draw [style=directed] (0) to (2.center);
		\draw [style=directed] (8.center) to (9.center);
	\end{pgfonlayer}
	\begin{pgfonlayer}{boxlayer}
		\draw [style=sep,fill=rouge] (-2.25, -0.75) rectangle (-0.25,2);
		\draw [style=sep,fill=bleu] (-1.75, 0.75) rectangle (-0.75, 1.75);
	\end{pgfonlayer}
	\begin{pgfonlayer}{framelayer}
		\path[fill=bleu] 
		($(current bounding box.south west) - (0.5,0)$) rectangle 
		($(current bounding box.north east) + (0.5,0.25)$); 
	\end{pgfonlayer}
\end{tikzpicture}}
  &\quad=\quad \vcenteredhbox{\begin{tikzpicture}
	\begin{pgfonlayer}{nodelayer}
		\node [style=none] (3) at (-1, 0.75) {};
		\node [style=rel] (4) at (-2.75, 0.75) {$S$};
		\node [style=none] (5) at (-2.75, 0.25) {};
		\node [style=rel] (6) at (-1.75, 0.75) {$M$};
		\node [style=none] (10) at (-1, -0.75) {};
		\node [style=none] (11) at (-1.75, -0.25) {};
		\node [style=none] (12) at (-2.75, -0.25) {};
		\node [style=none] (13) at (-1, -1.25) {};
	\end{pgfonlayer}
	\begin{pgfonlayer}{edgelayer}
		\draw [style=wire,in=90, out=90, looseness=2.00] (3.center) to (6);
		\draw [style=directed] (4) to (5.center);
		\draw [style=wire,bend left=90, looseness=1.25] (11.center) to (12.center);
		\draw [style=directed] (10.center) to (13.center);
		\draw [style=directed, in=90, out=-90] (6) to (11.center);
		\draw [style=directed] (12.center) to (5.center);
		\draw [style=directed] (10.center) to (3.center);
	\end{pgfonlayer}
	\begin{pgfonlayer}{boxlayer}
		\draw [style=sep,fill=rouge] (-0.75, -0.75) rectangle (-3.5, 1.5);
		\draw [style=sep,fill=bleu] (-2.25, 0.3) rectangle (-3.25,1.25);
		\draw [style=ghost,fill=bleu] (-2.0, 0.3) rectangle (-0.85,1.4);
	\end{pgfonlayer}
	\begin{pgfonlayer}{framelayer}
		\path[fill=bleu] 
		($(current bounding box.south west) - (0.5,0)$) rectangle 
		($(current bounding box.north east) + (0.5,0.25)$); 
	\end{pgfonlayer}
\end{tikzpicture}}
\end{align}
Besides capturing isomorphism, one can also express natural transformations between presheaves as certain diagrammatic moves or ``surgeries''.
The unit and counit of the two families of adjunctions
$$
R\otimes- \dashv R \multimap -\qquad
\pushD{M} \dashv \pullD{M}$$
yield directed versions of rule (\ref{equation/doublesep}),
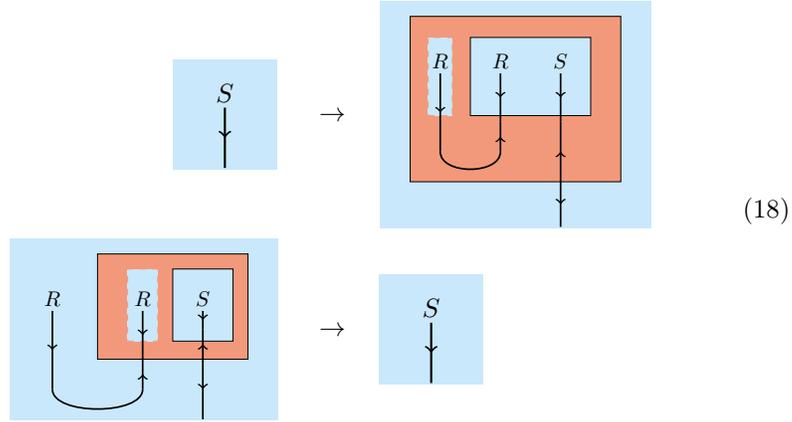
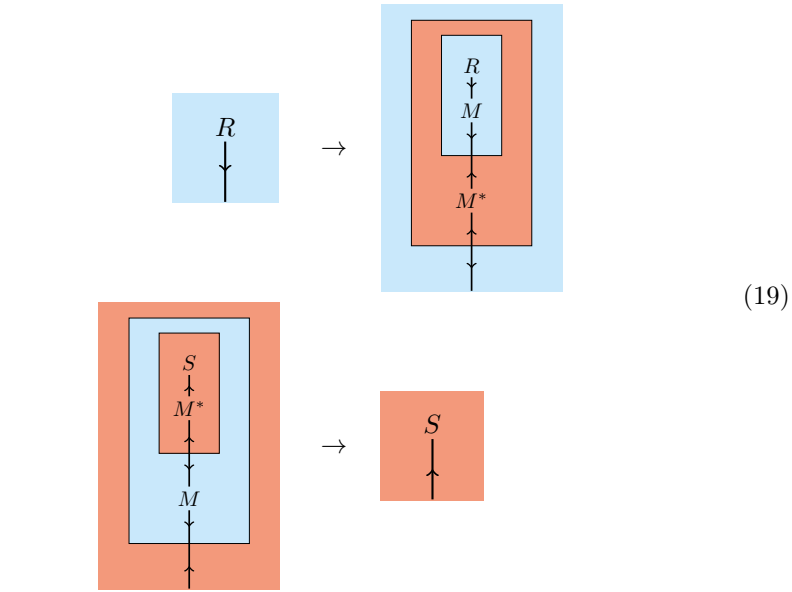
\begin{align}
\label{equation/evalcoeval}
\begin{split}
\vcenteredhbox{\scalebox{1}{\begin{tikzpicture}
	\begin{pgfonlayer}{nodelayer}
		\node [style=rel] (0) at (0.75, 1.75) {$S$};
		\node [style=none] (2) at (0.75, 0.75) {};
	\end{pgfonlayer}
	\begin{pgfonlayer}{edgelayer}
		\draw [style=directed] (0) to (2.center);
	\end{pgfonlayer}
	\begin{pgfonlayer}{framelayer}
		\path[fill=bleu] 
		($(current bounding box.south west) - (0.5,0)$) rectangle 
		($(current bounding box.north east) + (0.5,0.25)$); 
	\end{pgfonlayer}
\end{tikzpicture}}}
\quad&\to\quad
\vcenteredhbox{\scalebox{0.8}{\begin{tikzpicture}
	\begin{pgfonlayer}{nodelayer}
		\node [style=none] (1) at (2, 0.75) {};
		\node [style=none] (2) at (3, -0.25) {};
		\node [style=rel] (3) at (1, 1.75) {$R$};
		\node [style=rel] (4) at (3, 1.75) {$S$};
		\node [style=none] (8) at (2, 0.25) {};
		\node [style=none] (11) at (3, 0.75) {};
		\node [style=rel] (12) at (2, 1.75) {$R$};
		\node [style=none] (13) at (1, 0.25) {};
		\node [style=none] (19) at (3, -1) {};
	\end{pgfonlayer}
	\begin{pgfonlayer}{edgelayer}
		\draw [style=directed] (8.center) to (1.center);
		\draw [style=directed] (2.center) to (11.center);
		\draw [style=directed] (12) to (1.center);
		\draw [style=directed] (3) to (13.center);
		\draw [style=directed] (4) to (11.center);
		\draw [style=directed] (2.center) to (19.center);
		\draw [style=wire,bend right=90] (13.center) to (8.center);
	\end{pgfonlayer}
	\begin{pgfonlayer}{boxlayer}
		\draw [style=sep,fill=rouge] (0.5, -0.25) rectangle (4, 2.5);
		\draw [style=sep,fill=bleu] (1.5, 0.85) rectangle (3.5, 2.15);
		\draw [style=ghost,fill=bleu] (0.8, 0.85) rectangle (1.2,2.15);
	\end{pgfonlayer}
	\begin{pgfonlayer}{framelayer}
		\path[fill=bleu] 
		($(current bounding box.south west) - (0.5,0)$) rectangle 
		($(current bounding box.north east) + (0.5,0.25)$); 
	\end{pgfonlayer}
\end{tikzpicture}}} \\
\vcenteredhbox{\scalebox{0.8}{\begin{tikzpicture}
	\begin{pgfonlayer}{nodelayer}
		\node [style=none] (1) at (2.25, -0.25) {};
		\node [style=none] (2) at (2.25, 1.25) {};
		\node [style=rel] (7) at (2.25, 1.75) {$S$};
		\node [style=none] (9) at (1.25, 0.25) {};
		\node [style=none] (11) at (2.25, 0.75) {};
		\node [style=none] (12) at (1.25, 0.75) {};
		\node [style=rel] (13) at (1.25, 1.75) {$R$};
		\node [style=none] (15) at (-0.25, 0.25) {};
		\node [style=rel] (16) at (-0.25, 1.75) {$R$};
	\end{pgfonlayer}
	\begin{pgfonlayer}{edgelayer}
		\draw [style=directed] (7) to (2.center);
		\draw [style=directed] (11.center) to (2.center);
		\draw [style=directed] (9.center) to (12.center);
		\draw [style=directed] (11.center) to (1.center);
		\draw [style=directed] (13) to (12.center);
		\draw [style=directed] (16) to (15.center);
		\draw [style=wire,in=-90, out=-90, looseness=0.75] (15.center) to (9.center);
	\end{pgfonlayer}
	\begin{pgfonlayer}{boxlayer}
		\draw [style=sep,fill=rouge] (0.5, 0.75) rectangle (3, 2.5);
		\draw [style=sep,fill=bleu] (1.75, 1.05) rectangle (2.75, 2.25);
		\draw [style=ghost,fill=bleu] (1, 1.05) rectangle (1.5, 2.25);
	\end{pgfonlayer}
	\begin{pgfonlayer}{framelayer}
		\path[fill=bleu] 
		($(current bounding box.south west) - (0.5,0)$) rectangle 
		($(current bounding box.north east) + (0.5,0.25)$); 
	\end{pgfonlayer}
\end{tikzpicture}}}
\quad&\to\quad
\vcenteredhbox{\scalebox{1}{\begin{tikzpicture}
	\begin{pgfonlayer}{nodelayer}
		\node [style=rel] (0) at (0.75, 1.75) {$S$};
		\node [style=none] (2) at (0.75, 0.75) {};
	\end{pgfonlayer}
	\begin{pgfonlayer}{edgelayer}
		\draw [style=directed] (0) to (2.center);
	\end{pgfonlayer}
	\begin{pgfonlayer}{framelayer}
		\path[fill=bleu] 
		($(current bounding box.south west) - (0.5,0)$) rectangle 
		($(current bounding box.north east) + (0.5,0.25)$); 
	\end{pgfonlayer}
\end{tikzpicture}}}
\end{split}
\end{align}
\begin{align}
\label{equation/unitcounit}
\begin{split}
\vcenteredhbox{\scalebox{1}{\begin{tikzpicture}
	\begin{pgfonlayer}{nodelayer}
		\node [style=rel] (0) at (0.75, 1.75) {$R$};
		\node [style=none] (2) at (0.75, 0.75) {};
	\end{pgfonlayer}
	\begin{pgfonlayer}{edgelayer}
		\draw [style=directed] (0) to (2.center);
	\end{pgfonlayer}
	\begin{pgfonlayer}{framelayer}
		\path[fill=bleu] 
		($(current bounding box.south west) - (0.5,0)$) rectangle 
		($(current bounding box.north east) + (0.5,0.25)$); 
	\end{pgfonlayer}
\end{tikzpicture}}}
\quad&\to\quad
\vcenteredhbox{\scalebox{0.8}{\begin{tikzpicture}
	\begin{pgfonlayer}{nodelayer}
		\node [style=none] (0) at (5.5, -2.5) {};
		\node [style=rel] (2) at (5.5, -1) {$M^*$};
		\node [style=rel] (5) at (5.5, 0.5) {$M$};
		\node [style=none] (8) at (5.5, -0.25) {};
		\node [style=rel] (12) at (5.5, 1.25) {$R$};
		\node [style=none] (15) at (5.5, -1.75) {};
	\end{pgfonlayer}
	\begin{pgfonlayer}{edgelayer}
		\draw [style=directed] (12) to (5);
		\draw [style=directed] (2) to (8.center);
		\draw [style=directed] (5) to (8.center);
		\draw [style=directed] (15.center) to (0.center);
		\draw [style=directed] (15.center) to (2);
	\end{pgfonlayer}
	\begin{pgfonlayer}{boxlayer}
		\draw [style=sep,fill=rouge] (4.5, -1.75) rectangle (6.5, 2);
		\draw [style=sep,fill=bleu] (5, -0.25) rectangle (6,1.75);
	\end{pgfonlayer}
	\begin{pgfonlayer}{framelayer}
		\path[fill=bleu] 
		($(current bounding box.south west) - (0.5,0)$) rectangle 
		($(current bounding box.north east) + (0.5,0.25)$); 
	\end{pgfonlayer}
\end{tikzpicture}}}
\\
\vcenteredhbox{\scalebox{0.8}{\begin{tikzpicture}
	\begin{pgfonlayer}{nodelayer}
		\node [style=none] (0) at (5.5, -2.5) {};
		\node [style=rel] (2) at (5.5, -1) {$M$};
		\node [style=rel] (5) at (5.5, 0.5) {$M^*$};
		\node [style=none] (8) at (5.5, -0.25) {};
		\node [style=rel] (12) at (5.5, 1.25) {$S$};
		\node [style=none] (15) at (5.5, -1.75) {};
	\end{pgfonlayer}
	\begin{pgfonlayer}{edgelayer}
		\draw [style=directed] (5) to (12);
		\draw [style=directed] (8.center) to (2);
		\draw [style=directed] (8.center) to (5);
		\draw [style=directed] (0.center) to (15.center);
		\draw [style=directed] (2) to (15.center);
	\end{pgfonlayer}
	\begin{pgfonlayer}{boxlayer}
		\draw [style=sep,fill=bleu] (4.5, -1.75) rectangle (6.5, 2);
		\draw [style=sep,fill=rouge] (5, -0.25) rectangle (6,1.75);
	\end{pgfonlayer}
	\begin{pgfonlayer}{framelayer}
		\path[fill=rouge] 
		($(current bounding box.south west) - (0.5,0)$) rectangle 
		($(current bounding box.north east) + (0.5,0.25)$); 
	\end{pgfonlayer}
\end{tikzpicture}}}
\quad&\to\quad
\vcenteredhbox{\scalebox{1}{\begin{tikzpicture}
	\begin{pgfonlayer}{nodelayer}
		\node [style=rel] (0) at (0.75, 1.75) {$S$};
		\node [style=none] (2) at (0.75, 0.75) {};
	\end{pgfonlayer}
	\begin{pgfonlayer}{edgelayer}
		\draw [style=directed] (2.center) to (0);
	\end{pgfonlayer}
	\begin{pgfonlayer}{framelayer}
		\path[fill=rouge] 
		($(current bounding box.south west) - (0.5,0)$) rectangle 
		($(current bounding box.north east) + (0.5,0.25)$); 
	\end{pgfonlayer}
\end{tikzpicture}}}
\end{split}
\end{align}
where rule (\ref{equation/evalcoeval}) reduces to (\ref{equation/doublesep}) in the case that $R = 1$, and (\ref{equation/unitcounit}) to (\ref{equation/doublesep}) in the case that $M = \id$.

\section{Comparison with linear logic}\label{section/comparison-with-linear-logic}
One main benefit of our approach based on chiralities is that it enables us
to recover the dualities of classical logic in categorical situations
like the subset hyperdoctrine~$\predicates{}$ on~$\Set$,
or the presheaf hyperdoctrine~$\predicates{}$ on~$\Cat$, which are traditionally
seen as intuitionistic.
We thus find instructive to explain in what sense the six principles~(\ref{eqn:a})--(\ref{eqn:f}) asserted
for the monoidal closed bifibration chirality
$$(p,q):(\PSH,\coPSH)\to(\Dist,\Dist)$$
generalise well-known principles of linear logic.
To that purpose, we consider a $\ast$-autonomous category $\Vcategory$
with finite products and coproducts and we construct
two categories~$\Matplus(\Vcategory)$ and~$\Matminus(\Vcategory)$
whose objects~$A,B$ are the finite sets 
seen as discrete categories ; and whose morphisms $M:A\dto B$ 
are $\Vcategory$-valued matrices defined as functors $M:B\times A\to \Vcategory$.
Composition of $M:A\dto B$ and $N:B\dto C$ in $\Matplus(\Vcategory)$ is defined as
$$
N\circ M \quad = \quad (c,a) \quad \mapsto \quad \bigoplus_{b\in B} \, N(c,b)\otimes M(b,a).
$$
whereas composition of $M:A\dto B$ and $N:B\dto C$ in $\Matminus(\Vcategory)$ is defined as
$$
N\diamond M \quad = \quad (c,a) \quad \mapsto \quad \bigwith_{b\in B} \, N(c,b)\invamp M(b,a).
$$
$\MAT(\Vcategory)$ is the category whose objects are the $\Vcategory$-valued
matrices $R:1\dto A$, $S:1\dto B$, which may be alternatively seen as families
$\set{R_a\mid a\in A}$ or $\set{S_b\mid b\in B}$
of objects of~$\Vcategory$ ; and whose morphisms $R\to S$ are the pairs
$(M,\alpha)$ consisting of a matrix $M:A\dto B$ and of a natural transformation
$\alpha:M\circ R\Rightarrow S$, which may be alternatively seen
as a family of morphisms living in the category~$\Vcategory$
$$
\alpha_{a,b} \quad : \quad M(b,a) \otimes R_a \quad \morph{} \quad S_b
$$
indexed by the pairs $(a,b)\in A\times B$.
Similarly, $\coMAT(\Vcategory)$ is the category whose objects are matrices $R : A \dto 1$, $S : B \dto 1$~;
and whose morphisms $R \to S$ are the pairs $(M,\alpha)$ 
consisting of a matrix $M : A \dto B$ together with a natural transformation $\alpha :  R\Rightarrow S\diamond M$,
which may be alternatively seen as families of morphisms living in the category~$\Vcategory$
$$
\alpha_{a,b} \quad : \quad R_a \quad \morph{} \quad S_b\invamp M(b,a)
$$
indexed by the pairs $(a,b)\in A\times B$.
The forgetful functors $p:\MAT(\Vcategory)\to\Matplus(\Vcategory)$
and $q:\coMAT(\Vcategory)\to\Matminus(\Vcategory)$ define a bifibration chirality
$$
\xymatrix @-1.9pc {
\MAT(\Vcategory)\ar[ddddd]_-{p}
\ar@<.5ex>[rrrrrrrr]^-{(-)^{\ast}}
&&&&&&&&
\ar@<.5ex>[llllllll]^-{^{\ast}(-)}
\coMAT(\Vcategory)^{\,op}\ar[ddddd]^-{q^{\,op}}
\\
\\
\\
\\
\\
\Matplus(\Vcategory)
\ar@<.5ex>[rrrrrrrr]^-{(-)^{\ast}}
&&&&&&&&
\ar@<.5ex>[llllllll]^-{^{\ast}(-)}
\Matminus(\Vcategory)^{\,op}
}
$$
where a finite set~$A$ in $\MAT(\Vcategory)$ is transported by $(-)^*$ to itself: $A^{\ast}=A$ ;
where a morphism $M:A\dto B$ of $\Matplus$ is transported to the morphism $M^{\ast}:B\to A$ 
in $\Matminus$ obtained by flipping the inputs $A$ and $B$ and by applying pointwise negation
in $\Vcategory$:
$$
M^{\ast} \quad = \quad (a,b) \quad \mapsto \quad (M(b,a))^{\ast}
$$
and where the object $R^{\ast}: A\dto 1$ in $\coMAT$ is defined
by pointwise negation in $\Vcategory$:
$$
R^{\ast} \quad = \quad a \quad \mapsto \quad (R_a)^{\ast}
$$
The fact that $(p,q)$ define a bifibration chirality follows from the existence of the natural bijection 
$$
\Vcategory(M(b,a)\otimes R_a, S_b) \quad \cong \quad \Vcategory(S_b^*, R_a^*\invamp M^*(a,b))
$$
in the $\ast$-autonomous category~$\Vcategory$.
Given a $\Vcategory$-valued matrix $M:A\dto B$, 
the existential quantification of $R:1\dto A$ along~$M$
and the universal quantification of $R:1\dto B$ along $M:A\dto B$
in the bifibration $p$ are given by the formulas
$$
\exists_M \, R \quad = \quad b \quad \mapsto \quad \bigoplus_{a\in A} \,\, M(b,a)  \otimes  R_a
$$
\vspace{-1em}
$$
\forall_M \, S \quad = \quad  a \quad \mapsto \quad \bigwith_{b\in B} \,\, M^*(a,b)\invamp S_b
$$
The bifibration chirality $(p,q)$ is also monoidal closed, with
conjunction and disjunction defined pointwise:
$$
\begin{array}{ccccc}
R\tensorialand S &  =  & (a,b) \quad \mapsto \quad R_a\otimes S_b & \quad\quad \refs & A\times B
\\
R\tensorialor S &  =  & (a,b) \quad \mapsto \quad R_a\invamp S_b &\quad\quad \refs & A\times B
\end{array}
$$
where $R\refs A$ and $S\refs B$.
In this specific monoidal closed bifibration chirality~$(p,q)$,
the formulas (\ref{eqn:a})--(\ref{eqn:f}) enable us to recover familiar
principles of linear logic:
\begin{tcolorbox}[left=1mm]
$$\begin{array}{ccc}
\\
 & & \with_{b\in B} ( S_b \invamp M(b,a) ) 
\\
(a,b) & \!\! \viso \!\! & (\,\, \oplus_{b\in B}\, M^*(a,b) \otimes S_b^* \,\,)^*
\\
\\
 & & \big( \oplus_{a\in A} M(c,a)\otimes R_a \big) \otimes \big( \oplus_{b\in B} N(d,b)\otimes S_{b} \big)
\\
(c) & \!\! \viso \!\! &
\oplus_{(a,b)\in A\times B} \big( M(c,a) \otimes N(d,b)  \otimes Ra \otimes S_b\big)
\\
\\
 & & \big( \with_{a\in A} R_a \invamp M(a,c) \big) \invamp \big( \with_{b\in B} S_b \invamp N(b,d) \big)
\\
(d) & \!\! \viso \!\! &
\with_{(a,b)\in A\times B} \big( R_a \invamp S_b \invamp M(a,c) \invamp N(b,d)  \big)
\\
\\
& & \big( \with_{a\in A} R_c \invamp M(c,a)\big) \otimes \big( \with_{b\in B} S_d \invamp N(d,b) \big)
\\
(e)  & \!\! \to \!\! &
\with_{(a,b)\in A\times B} \big( \big( R_a\otimes S_b \big) \invamp N(d,b) \invamp M(c,a)\big)
\\
\\
& & \oplus_{(a,b)\in A\times B}  
\big(  N(d,b)\otimes M(c,a) \otimes (R_c\invamp S_d) \big)
\\
(f)  & \!\! \to \!\! &
\big( \oplus_{a\in A} M(c,a) \otimes R_a \big) \invamp \big( \oplus_{b\in B} N(d,b)\otimes S_b\big)
\\
\\
\end{array}$$
\end{tcolorbox}

\noindent {\bf Acknowledgments.} 
This work has received funding from the European Research Council (ERC) under the ERC Advanced Grant DuaLL (grant agreement No 670624) and the ERC Advanced Grant ProofCert.

\medbreak

\end{document}